\documentclass[epj]{svjour}
\usepackage{graphics}
\usepackage{graphicx}
\usepackage{epsfig,hyperref}
\usepackage{ulem}
\usepackage{morefloats}
\usepackage{rotating}
\usepackage{color,xcolor}
\usepackage{soul}
\usepackage{amsmath}
\usepackage{amssymb}
\usepackage{url}
\usepackage[utf8]{inputenc}
\usepackage{relsize}

\usepackage{bm} 
\newcommand{\bwt}{\begin{widetext}}
\newcommand{\ewt}{\end{widetext}}
\newcommand{\beq}{\begin{equation}}
\newcommand{\eeq}{\end{equation}}
\newcommand{\bea}{\begin{eqnarray}}
\newcommand{\eea}{\end{eqnarray}}

\begin{document}

\title{ Crust-core transition of a neutron star: effect of the 
temperature under strong magnetic fields}

\author{Márcio Ferreira\inst{1},  Aziz Rabhi\inst{1}, Constan\c ca Provid{\^e}ncia\inst{1}}

\mail{Constan\c ca Provid\^encia, \texttt{cp@uc.pt}}

\institute{CFisUC, Department of Physics, University of Coimbra,
   3004-516 Coimbra, Portugal.}

\date{Received: date / Revised version: date}

\abstract{
The effect of  temperature on the crust-core
transition of a magnetar is studied. The
thermodynamical spinodals are used to calculate the
transition region within a relativistic mean-field approach for the
equation of state. Magnetic fields with intensities $5\times 10^{16}$ G
and $5\times 10 ^{17}$G
are considered. It is shown that the
effect on the extension of the crust-core transition is washed away for
temperatures above $10^{9}$ K for magnetic field intensities $
\lesssim 5\times 10^{16}$ G but may still persist if  a magnetic field
as high as $5\times 10 ^{17}$G is considered. For  temperatures below that value, the
effect  of the magnetic field on crust-core transition is noticeable
and  grows as the temperature decreases and, in particular,
it is interesting to identify the existence of disconnected
non-homogenous matter above the $B=0$ crust core transition density.
Models with different symmetry energy slopes at saturation show quite
different behaviors. In particular, a model with a large slope, as
suggested by the recent results of PREX-2, predicts the existence of up
to four disconnected regions of non-homogeneous matter above the zero
magnetic field crust-core transition density.
}

\maketitle

\section{Introduction}

Magnetars are considered to be highly magnetized young neutron stars which
present  a
very wide  variety  of different types of  activity, and, in
particular, are sources of huge amounts of energy in the form of both  electromagnetic  and  gravitational
waves, see
\cite{Kaspi2017} for a review. In particular,
soft-gamma repeaters (SGRs) and anomalous x-ray pulsars
(AXPs) were some of the first  magnetars to be identified in
astronomy  \cite{Duncan1992,Paczynski1992}.   These neutron stars have
surface magnetic fields as high as $10^{14}-10^{15}$ G. Presently
about 30 neutrons stars  are known that fall in this category \cite{kaspi2014}\footnote{SGR/APX online catalogue,
       url{http://www.physics.mcgill.ca/~pulsar/magnetar/main.html}}.

In \cite{Fang16,Fang2017,Fang2017a,Avancini2018}, the authors have discussed, using a dynamical spinodal approach that the extension of the spinodal region
of nuclear matter in the space defined by the proton and neutron
densities, i.e. ($\rho_p,\, \rho_n$),  is larger  than the $B=0$  spinodal
region for magnetic field intensities larger than
10$^{16}$ G. Moreover, they have predicted from the crossing of  a
well defined proton fraction curve, as expected inside neutron stars,
and the spinodal surface that extra disconnected non-homogeneous
regions could exist close but at larger densities than the $B=0$
crust-core transition. To determine the dynamical spinodal, the region
where the frequency of density
fluctuations goes to zero was calculated. A similar conclusion was drawn under a
thermodynamical spinodal approach in
\cite{Rabhi2008a,Rabhi2009a,Chen17,Fang2017a,Chatterjee2018}.  In this approach the
instability region corresponds to the region
where the curvature of the free energy density is negative. In
\cite{Avancini2010}, it has been shown that both approaches give a good
prevision of the crust-core transition density inside a neutron star,
although the prediction obtained with the thermodynamical spinodal
could be approximately 10\% larger.

The main reason for the
different behavior  of magnetized with respect to non-magnetized
nuclear matter is dictated by the Landau quantization of the proton energy
due to its charge as discussed in \cite{Broderick2000}. Besides,
including the anomalous magnetic moment (AMM) of both nucleons an
extra polarization effect is expected that also affects the properties
of nuclear matter in thermodynamic equilibrium \cite{Broderick2000,PerezGarcia2011}.
In the neutron star core very strong magnetic fields are necessary to
affect the EoS, see for instance
\cite{Broderick2000,Broderick02,Rabhi2008}, while from the
virial  theorem  \cite{Lai1991} or from the integration of the Einstein equations coupled
to the electromagnetic ones \cite{Bocquet1995,Cardall2000,Chatterjee15} the
strongest fields that a neutron star can support are of the order of
10$^{18}$ G. Although, stronger fields are needed to affect the
structure of the core EoS, much weaker magnetic fields have a very
strong effect on the outer  neutron star crust and
the onset of the neutron drip as shown in \cite{Chamel2012,Chamel2015}
(see also \cite{Blaschke2018}). Also the inner crust is affected by
fields of the order of 10$^{17}-10^{18}$ G as shown in
\cite{Lima-13,Bao21}, and more recently even with weaker fields
\cite{Pais2021}. In \cite{Sengo20}, it was shown that the neutral 
magnetic field line of a poloidal configuration could fall inside an extended crust
as predicted in \cite{Fang16,Fang2017,Pais2021}, which could explain
effects as the fracture of the crust.

A neutron star is formed with temperatures above $10^{10}$ K but
within minutes its temperature is below $10^{9}$ K, and after several kyrs
it is  expected to have cooled down to
surface temperatures below 10$^6$ K  and an inner temperature of the
order of 10$^8$ K \cite{Yakovlev2004}. However, in \cite{Vigano2013} it
was shown that young  magnetars with strong surface magnetic fields $\sim
10^{15}$ G could have surface temperatures 
magnetars up to $5\times 10^6$ K and inner temperatures,  at the bottom
of the envelope corresponding to  a density of the order of $10^{10}- 3\times 10^{11}$g/cm$^3$ , up to $\sim 5\times
10^{8}$ K. Similar conclusions were drawn in other studies \cite{Potekhin2001,Pons2009,Potekhin2018,Akgun2018}.
According to \cite{Duncan1992,Paczynski1992} the large
luminosities of magnetars originate on the huge amounts of magnetic
energy those stars store.
In \cite{Beloborodov2016} the authors have
studied several mechanisms, such as mechanical dissipation
in the solid crust or heat flux from the  liquid core, that convert
the magnetic field energy into heat.

However, 
it is worth mentioning a recent study on the composition of the outer crust of non-accreting neutron star  where it has been discussed that the composition of the cristalized crust is not the one  determined by the cold catalyzed-matter  hypothesis but the one occuring at the  crystallization temperature of the crust. This scenario is expected if the NS cooling  is fast compared to the nuclear reaction rates that determine cold catalyzed-matter \cite{Arnould2007,Goriely2011}.  Although in \cite{Carreau2020} no magnetic field effects were considered,  it raises extra questions concerning the composition of the magnetar crust that need to be addressed in the future.

The possibility that magnetars
have temperatures in their interior larger than the ones expected
inside a neutron star sets the question whether temperature may wash away
the effect of Landau quantization, and in particular,  the enlarged
extension of the crust predicted at $T=0$ MeV. A first study was
undertaken in \cite{Fang2017a}, where the effect of temperatures below
1 MeV on the crust properties of the magnetized stars with crust
magnetic fields equal to  $5\times 10^{15}- 5\times 10^{16}$  was
analysed. It was shown that a temperature of the order of 10 keV would
wash away the effect of the enlargement of the crust for fields below
$2\times 10^{16}$ G.  In the present study  we want to deepen the
analysis of the competition between the effect of Landau quantization
and   temperature on the crust-core transition. However,  we only consider temperature effects on the single-particle occupation of excited states above the Fermi surface. Other temperature effects may affect the structure of the crust, in particular, the contribution of the  electron-ion interaction to the melting of Coulomb  lattice as discussed in \cite{Potekhin2000}, or considering that the pasta phases behave as  liquid crystals instead of  rigid solids \cite{Pethick1998} the thermally induced displacements and deformation of the clusters may lead to the melting of the Coulomb lattice at very low temperatures \cite{Watanabe2000a,Watanabe2000b}.

The present paper is organised in the following way: in
Sec. \label{sec2} we make a brief presentation of the  nuclear models that
will be applied in the analysis; in Sec. \ref{sec3} we review the
formalism used to calculate the  thermodynamic spinodals; in Sec. \ref{sec4} the results for
two magnetic field intensities and several temperatures  are presented and
discussed and in Sec. \ref{sec5} some conclusions are drawn.

\section{Nuclear matter model\label{sec2}}

In the present section we review the formalism applied to described
magnetized stellar matter within a nuclear RMF approach
\cite{Broderick2000,Rabhi08}. We also introduce the 
anomalous magnetic moment
(AMM) and will discuss its effect.
We consider the usual  Lagrangian density  
$$
{\cal L}=\sum_{i=p,n} {\cal L}_i + {\cal L}_e + \cal L_\sigma + {\cal
  L}_\omega + {\cal L}_\rho + \mathcal{L}_{\omega \rho },$$
where  the different terms are: ${\cal L}_i$ the nucleon Lagrangian density, given by
\begin{equation}
{\cal L}_i=\bar \psi_i\left[\gamma_\mu i
  D^\mu-M^*_i-\frac{1}{2}\mu_N\kappa_i\sigma_{\mu \nu} F^{\mu
    \nu}\right]\psi_i,
\label{lagr}
\end{equation}
with
$$
iD^\mu=i \partial^\mu-g_v V^\mu-
\frac{g_\rho}{2}\boldsymbol\tau \cdot \mathbf{b}^\mu - e A^\mu
\frac{1+\tau_3}{2}\,,
$$
and
$
M^*_p=M^*_n=M^*=m-g_s\phi;
$
 the mesonic terms  for the $\sigma$, $\omega$ and $\rho$-mesons,  $\cal L_\sigma$,  ${\cal
   L}_\omega$ and ${\cal L}_\rho$, are  defined as in \cite{Fang2017} in
 terms of the fields $\phi$, $V^\mu$ and $\boldsymbol{b}^\mu$ with
 masses $m_\sigma$, $m_v$ and $m_\rho$, respectively,
 \begin{eqnarray}
{\cal L}_\sigma&=&\frac{1}{2}\left(\partial_\mu\phi\partial^\mu\phi-m_\sigma^2 \phi^2 - \frac{1}{3}\kappa \phi^3 -\frac{1}{12}\lambda\phi^4\right),\nonumber\\
{\cal L}_\omega&=&-\frac{1}{4}\Omega_{\mu\nu}\Omega^{\mu\nu}+\frac{1}{2}
m_v^2 V_\mu V^\mu+\frac{1}{4!} \xi g_v^4 \left(V_\mu V^\mu\right)^2, \nonumber \\
{\cal L}_\rho&=&-\frac{1}{4}\mathbf B_{\mu\nu}\cdot\mathbf B^{\mu\nu}+\frac{1}{2}
m_\rho^2 \mathbf b_\mu\cdot \mathbf b^\mu,
\end{eqnarray}
where
$\Omega_{\mu\nu}=\partial_\mu V_\nu-\partial_\nu V_\mu$, and $\mathbf B_{\mu\nu}=\partial_\mu\mathbf b_\nu-\partial_\nu \mathbf b_\mu
- g_\rho (\mathbf b_\mu \times \mathbf b_\nu)$, and the parameters
$\kappa$ and $\lambda$ are associated with the third- and fourth-order terms of the scalar field.  
The term
$$\mathcal{L}_{\omega \rho } = \Lambda_v g_v^2 g_\rho^2 V_{\mu }V^{\mu }
\mathbf{b}_{\mu }\cdot \mathbf{b}^{\mu },$$
couples the $\rho$  and the
$\omega$ meson.  Changing the strength of this term will allow to make
the symmetry energy softer (harder) increasing (decreasing) the
coupling  \cite{Horowitz2001PRL,Providencia2013,Pais2016Vlasov}. In our study we
consider the models TM1 \cite{tm1} and  TM1e \cite{tm1e}. Both
parametrizations describe symmetric nuclear matter with the
same functional and only differ in the isovector  channel. Both describe  two solar
mass stars \cite{fortin16,tm1e,Shen2020}.
Finally, ${\cal L}_e$ is the electron Lagrangian density,
Electrons, with mass $m_e$, are also included in the Lagrangian density,
$${\cal L}_e=\bar \psi_e\left[i\gamma_\mu
  \partial^\mu-m_e\right]\psi_e,$$
when the $\beta$-equilibrium EoS is calculated. Since we will only
discuss the crust-core transition, muons will not be included as they
set in at $\rho\gtrsim 0.11$ fm$^{-3}$ above the crust-core transition.

Protons interact with a static  field
$A^{\mu}$, which  includes a static component  assumed to be
externally generated, $A^{\mu}=(0,0,Bx,0)$, 
so that $\mathbf{B}=B\, \hat{z}$ and $\nabla \cdot {\bf A}$=0. 
The nucleon AMM is introduced via the coupling of the baryons to the
electromagnetic field tensor with $\sigma_{\mu\nu}=\frac{i}{2}\left[\gamma_{\mu}, \gamma_{\nu}\right] $,
and strength $\kappa_{i}$, with $\kappa_{n}=-1.91315$ for the neutron,
and $\kappa_{p}=1.79285$ for the proton. $\mu_N$ is the nuclear
magneton.

The properties of TM1 and TM1e models at saturation \cite{Shen2020} are the same for the
isoscalar channel,  in
particular, the binding energy  $E_b=-16.3$ MeV,  the saturation
density $\rho_0=0.145$ fm$^{-3}$, and the incompressibility $K= 281$
MeV.  The two models differ in the
isovector properties: the symmetry energy  $J$ and its
slope $L$ at saturation, take, respectively, the values 36.89 MeV and
110.8 MeV for TM1 and  31.38 MeV and 40 MeV for TM1e.
The last model satisfies the constraints set by 
microscopic calculations of neutron matter in a chiral effective field
approach \cite{Hebeler2013}.  On the other hand, recent results from
the Lead Radius EXperiment
(PREX-2) collaboration have  extracted from the measurements of the
parity-violating asymmetry a  neutron skin thickness
$R_n-R_p=0.283\pm0.071$ fm\cite{prex2}, which, according to
\cite{Reed2021}, corresponds to a slope of the symmetry energy at
saturation in the range $L=106\pm 37$ MeV, taking into account the
strong correlation existing between the neutron skin thickness and the
symmetry energy slope $L$.  These results justify that we also
consider  model TM1 with  larger symmetry energy than TM1e, well
inside the range proposed in \cite{Reed2021}.

\section{Thermodynamical  spinodal \label{sec3}}
In the present section we introduce the thermodynamical spinodal which
will be used to estimate the transition from a clusterized to
homogeneous region. In particular, we will show how the spinodal
changes with temperature and how temperature affects the transition
region.

The state which minimizes the free energy of asymmetric $npe$ matter
is characterized by the  nucleon distribution functions
\begin{equation}
f_{i \pm}= \frac{1}{1+e^{(\epsilon_{i} \mp \nu_i)/T}} \quad i=p,n,
\end{equation}
where the effective chemical potential $\nu_i$ is related to the
chemical potential $\mu_i$ by the equation
\begin{equation}
\nu_i=\mu_i - g_v V_0  - \frac{g_\rho}{2}\, \tau_i b_0 
\end{equation}
and the electron  distribution function
\begin{equation}
f_{e \pm}= \frac{1}{1+e^{(\epsilon_{e} \mp \mu_e)/T}},
\end{equation}
with  $\mu_e$  the electron chemical potential.
In these expressions the quasiparticle energies are given by
\begin{eqnarray}
\epsilon_p&=&\sqrt{{p_z}^2+\bar m_p^2} \nonumber\;,\\
\epsilon_n&=&\sqrt{{p_z}^2+\left(\sqrt{M^{* 2}+{p_\perp}^2}-s\mu_N\kappa_n B \right)^2} \nonumber\;,\\
\epsilon_e&=&\sqrt{ p_z^2+\bar m_e^2} \nonumber\;.
\end{eqnarray}
where $$\bar m_p=\tilde m_{p,\nu}-s\mu_N\kappa_p B\equiv \bar m_{p,\nu
  s}$$ with
$\tilde m_{p,\nu}= \sqrt{M^{* 2}+2\nu q_{p}B}$,  $ \bar m_{e}= \sqrt{m_e+2\nu e B} \equiv \bar
m_{e,\nu}$, $\nu$ designates the Landau level and $s=\pm1$  is the spin
projection in the direction of the magnetic field.

In the above expressions the  constant mesonic fields satisfy the following equations
$$ m_\sigma^2\phi_0 + \frac{\kappa}{2} \phi_0^2 + \frac{\lambda}{6} \phi_0^3 =g_s(\rho_{sp}+\rho_{sn}),$$
$$m_v^2\,V_0\,=\, g_v (\rho_{p}+\rho_{n}),$$
$$m_{\rho}^2\,b_0\,=\, \frac{g_\rho}{2}
(\rho_{p}-\rho_{n}),$$
the spatial components of the vector mesons being zero since we
consider a static description.
At zero temperature the above distribution functions reduce to
$f_{i+}=\theta(P_{Fi}^2-p^2)$ where $P_{Fi}$ is the Fermi momentum of
nucleon $i$  and  $f_{i-}= 0$ \cite{Brito06,Pais-10}.

In terms of the finite temperature distribution functions  the scalar and vector densities are given by
\begin{equation}
\rho_{sp}=\frac{eB}{(2\pi)^2}\sum_{\nu,s}\int  dp_z\,\left(f_{p+}+f_{p-}\right)\frac{\bar m_p M^*}{(\bar m_p+s\mu_N\kappa_p B)\epsilon_p} \;,
\end{equation}
\begin{equation}
\rho_{sn}=\sum_{s}\int  \frac{d^3p}{(2\pi)^3}\left
  (f_{n+}+f_{n-}\right)\left[1-\frac{s\mu_N\kappa_n B}{\sqrt{M^{*
        2}+p^2_\perp}}\right]
\frac{M^*}{\epsilon_n}\;,
\end{equation}
\begin{equation}
\rho_p=\frac{eB}{(2\pi)^2}\sum_{\nu,s}\int dp_z\, \left (f_{p+}-f_{p-}\right) \nonumber \;,
\end{equation}
\begin{equation}
\rho_n=\sum_{s}\int  \frac{d^3p }{(2\pi)^3} \left (f_{n+}-f_{n-}\right)\nonumber\;,
\end{equation}
\begin{equation}
 \rho_e=\frac{eB}{(2\pi)^2}\sum_{\nu,s}\int dp_z\,\left (f_{e+}-f_{e-}\right)\;.
 \end{equation}

The stability of nuclear matter is determined imposing that the free
energy density $\cal F$ is a convex function of the densities $\rho_p$
and $\rho_n$. For stable homogeneous matter, the symmetric curvature matrix with the elements \cite{Muller-95,Baran-98,Margueron2003},
\begin{equation}
{\cal F}_{ij}=\left( \frac{\partial^{2} {\cal F}}{\partial \rho_{i}\partial\rho_{j}}\right) _{T},
\end{equation}
must be positive. This is equivalent to imposing that the trace and the determinant of ${\cal F}_{ij}$ are positive.
In terms of the proton and neutron chemical potentials $\mu_i$, the curvature matrix is given by
\begin{equation}
{\cal F}=
\begin{pmatrix}
\displaystyle\frac{\partial\mu_{n}}{\partial \rho_{n}}
&\displaystyle\frac{\partial\mu_{n}}{\partial \rho_{p}} \\
\displaystyle\frac{\partial\mu_{p}}{\partial \rho_{n}}
&\displaystyle\frac{\partial\mu_{p}}{\partial \rho_{p}}
\label{cur}
\end{pmatrix}
\end{equation}
with $\mu_{i}=\frac{\partial{\cal F}}{\partial \rho_{i}}|_{T,\rho_{j\neq i}} $.
For nuclear matter, the largest eigenvalue of the curvature matrix is always positive and the other becomes negative at sub-saturation densities.
The surface characterized by a zero  determinant of $ {\cal
  F}_{ij}$, e.g.  a zero eigenvalue, defines the 
thermodynamical spinodal on the  ($\rho_n$,
$\rho_p$,$T$) space. Inside the thermodynamical spinodal the smallest
eigenvalue of  ${\cal F}_{ij}$ nuclear matter is negative
corresponding to an unstable state.
The curvature matrix has the eigenvalues
\begin{equation}
\lambda_{\pm}=\frac{1}{2}\left(\hbox{Tr}({\cal F})\pm\sqrt{\hbox{Tr}({\cal F})^2-4 \hbox{Det}({\cal F})}\right).
\label{lambd}
\end{equation}
According to the stability condition both must be positive. A
thermodynamic instablility  is obtained if one becomes negative. In
this case, the system will decrease its free energy evolving in the
instability direction, which is the direction of the eigenvector associated to the negative eigenvalue.
The eigenvectors $\delta {\bf \rho}_{\pm}$ are given by
\begin{equation}
\frac{\delta \rho^{\pm}_{i}}{\delta \rho^{\pm}_{j}}=\frac{\lambda_{\pm}-{\cal F}_{jj}}{{\cal F}_{ji}}, \: i, j = p, n.
\end{equation}

\begin{figure*}[!t]
  \begin{tabular}{cc}
\includegraphics[width=0.4\linewidth]{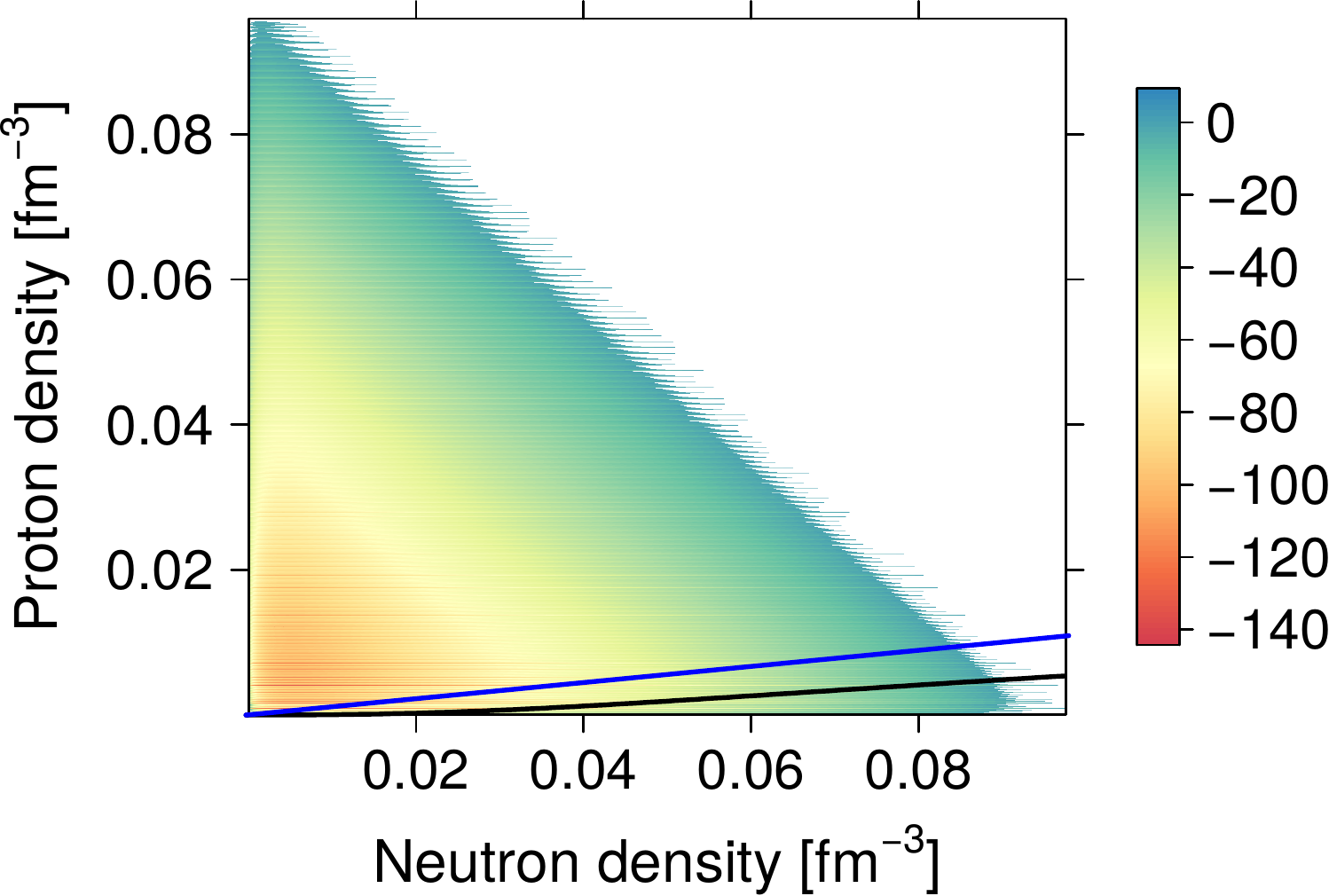} &

\includegraphics[width=0.4\linewidth]{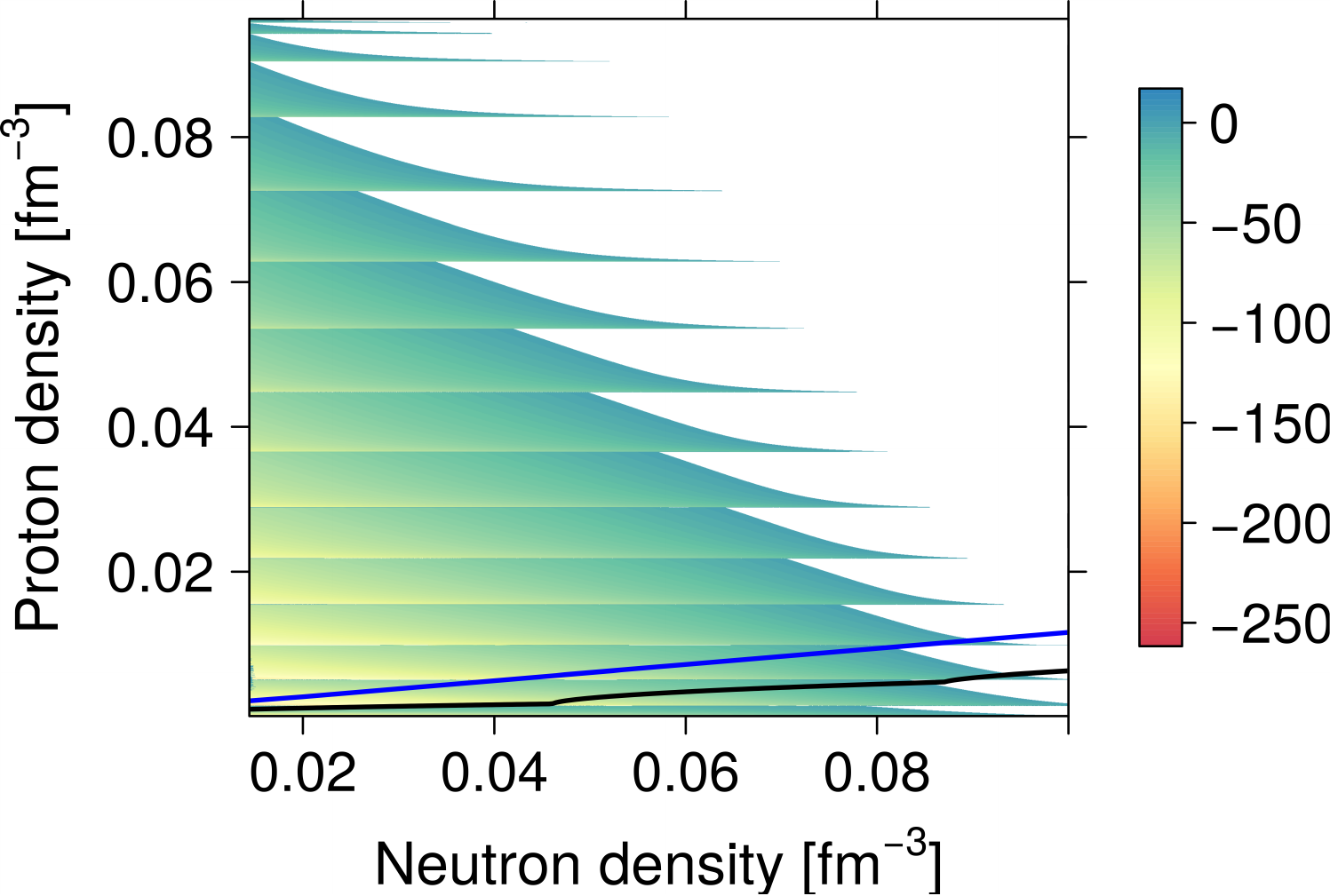}\\

\includegraphics[width=0.4\linewidth]{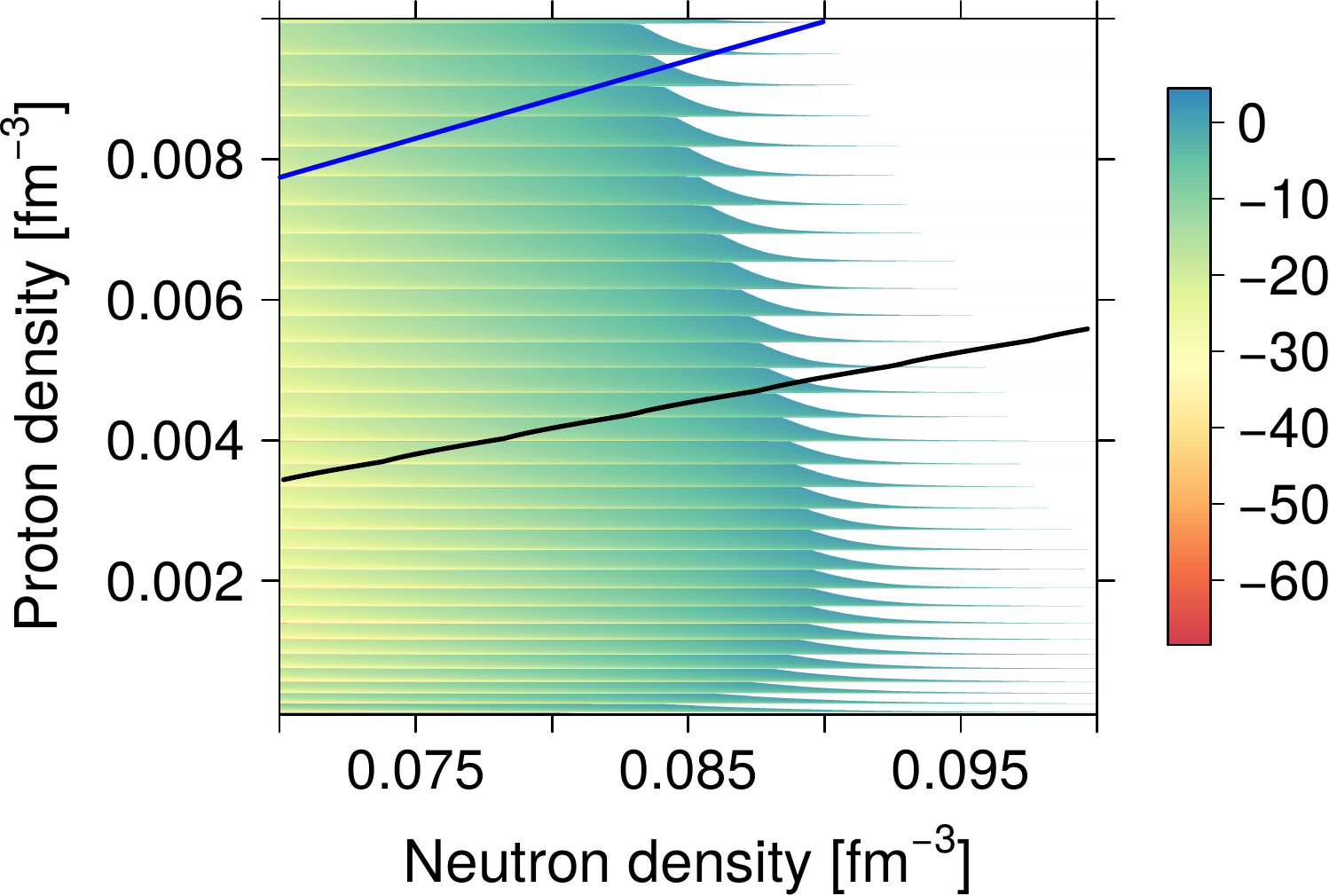} &

\includegraphics[width=0.4\linewidth]{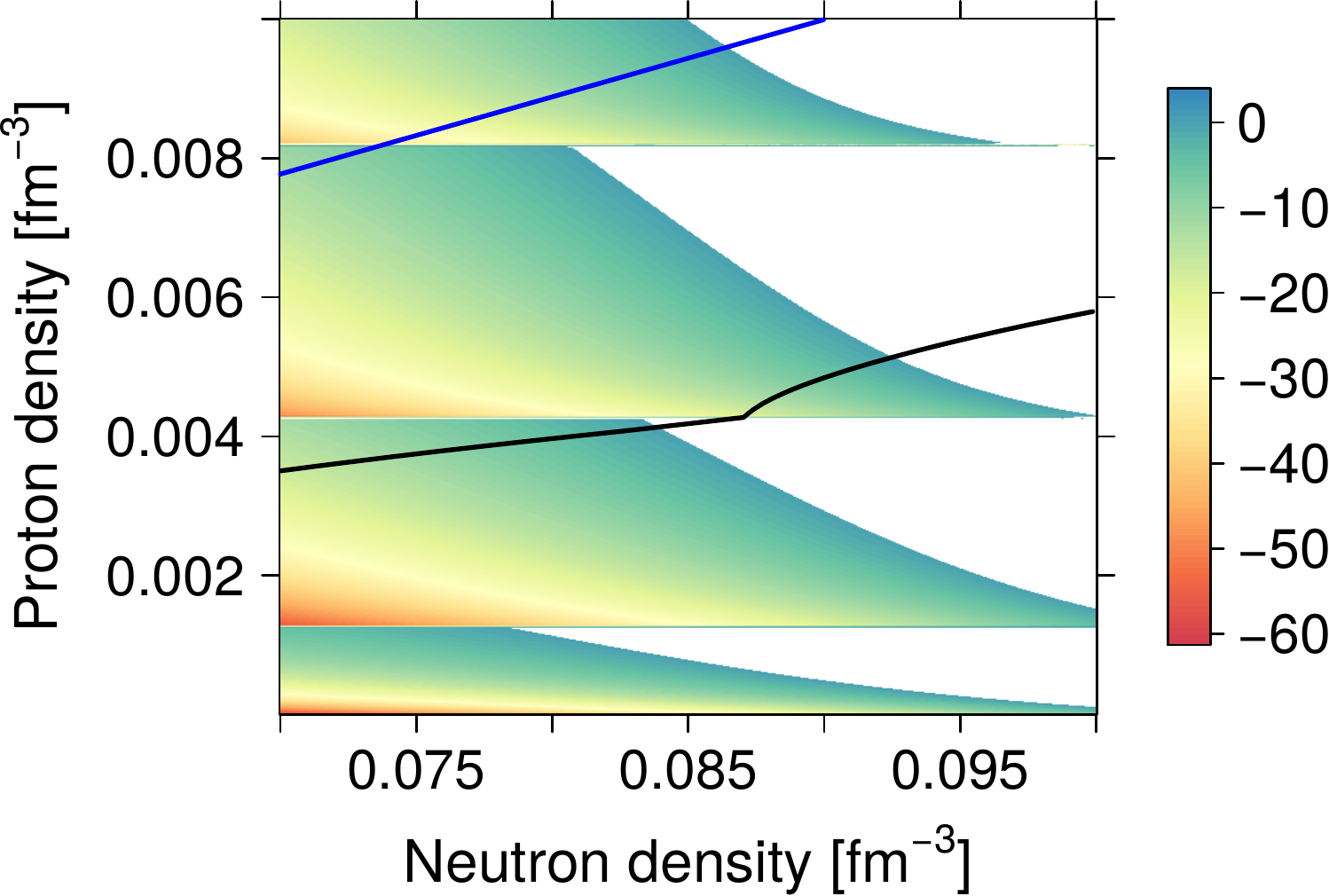}
   \end{tabular}                                                                           
\caption{Spinodal section obtained within the TM1e model 
for $B^{*}=10^{3}$ (left) and $B^{*}=10^{4}$ (right) at zero
temperature. The bottom panels contain a detail at the region of
interest for the crust-core transition.  No AMM has been considered. The colours show the
magnitude of $\lambda_-$ (see Eq. (\ref{lambd}) in units of $(\hbar
c)^3/M^2$. The two lines represented define the $Y_p=0.1$ (top) and
$\beta$-equilibrium (bottom) proton fractions.}
\label{T0}
\end{figure*}

The free energy density is given by
\beq
{\cal F}=\varepsilon -T{\cal S}
\eeq
where $T$ is the temperature and $\varepsilon$ is the  energy density 
\bea
\label{ener}
\varepsilon&=&\sum_{i=p,n} \varepsilon_{i}+\frac{1}
{2}m^{2}_{\sigma}\phi^{2}+ \frac{1}{3}\kappa \phi^3
+\frac{1}{12}\lambda\phi^4
\\
&&+
\frac{1}{2}m^{2}_{v} V^{2}_{0}+\frac{1}{2}m^{2}_{\rho} b^{2}_{0}
+\frac{1}{8}\xi g_{v}^4 V^{4}_{0}+ 3\Lambda_v  g_{v}^2
g_{\rho}^2 V^{2}_{0} b^{2}_{0}, \nonumber
\eea
and ${\cal S} = {\cal S}_p+ {\cal S}_n$  is the  entropy density. 
In the above expressions the  proton energy density  $\varepsilon_p$ and the neutron energy density
$\varepsilon_n$ are defined by \cite{Rabhi08}
\bea
\varepsilon_p&=&\frac{eB}{2\pi^ {2}}\sum_{\nu, s}\int^{\infty}_{0}
dp_{z}  \epsilon_p\left(
f_{p+}+{f}_{n-}\right) ,\\
\varepsilon_n&=&\sum_{s}\int \frac{d^3p}{(2\pi)^{3}}
\epsilon_n\left( f_{n+}+f_{n-}\right) 
\eea
and the proton and neutron  entropy densities by
\bea
{\cal S}_p&=&-\frac{e B}{2\pi^ {2}}\sum_{\nu, s}\int^{\infty}_{0}
dp_z\left\lbrace f_{p+}\ln f_{p+} \right. \\
&&\left.+(1-f_{p+})\ln(1- 
f_{p+})+  (p+ \to p-)
\right\rbrace  \nonumber\\
{\cal S}_n&=& -\sum_{s}\int \frac{dp^{3}}{(2\pi)^ {3}}
\left\lbrace  f_{n+}\log f_{n+}\right. \\
&&\left.+(1-f_{n+})\log(1- 
f_{n+})+  (n+ \to n-)
\right\rbrace. \nonumber
\eea

In order to calculate the crust-core transition, we will determine the
crossing of the stellar matter equation of state with the spinodal
surface. Two different scenarios will be considered: i)
$\beta$-equilibrium neutrino free cold and warm matter; ii) matter
with a charge fraction $Y_p=0.1$ a fraction that is larger than the
one of $\beta$-equilibrium and which occurs in the proto-neutron star
already after the diffusion of neutrinos out of the star. 
The crust-core transition will be determined from the crossing of the
$\beta$-equilibrium proton fraction, and the $Y_p=0.1$ curve with the
spinodal section. For $T=0$, it was shown that there may exist a
sequence of unstable bands separated by stable bands due to the
existence of Landau levels. However, the effects Landau quantization
are diluted by the inclusion of temperature and this effect will be
discussed in the following.

\section{Results \label{sec4}}

In the present section, we will show some calculations of the spinodal
surface performed for two reference magnetic field intensities:
$B^*=10^3$ and $10^4$ corresponding to 4.4 $\times 10^{16}$ G and 4.4 $\times 10^{17}$ G  since $B^*=B/B_{ce}$ with $B_{ce}=4.4\times 10^{13}$ G the
critical electron field. 
 These intensities  could occur in the interior of a magnetar
  \cite{Frieben2012,Lander2021,Uryu2019}. In
  \cite{Lander2021,Uryu2019}, and contrary to \cite{Frieben2012}, both
  a poloidal and toroidal component are considered, and  the
  localization of the toroidal lines of fields with a  magnitude  that
  can be as high as  $B^*=10^3$ to $10^4$ close to the surface, where
  one may expect the crust-core transition to occur. It should,
  however, be pointed out that in  \cite{Uryu2019} taking a single
  segment polytropic EOS, the huge fields were  considered, and it was
  shown that they affect the distribution of matter  and create a magnetovacuum region.  This indicates that further studies that couple the magnetic field to matter are required. 

The
magnetized matter will be calculated within the TM1 and TM1e models
\cite{tm1,tm1e} as referred before. These are models that differ in
the density dependence of the symmetry energy. We will consider the
temperatures $T=10,\, 50,\, 100$ and 500 keV. 
In the low magnetic field limit, we expect the energy difference
between Landau levels of the order $\Delta \epsilon\lesssim
eB/M^*=m_e^2\, B^*/M^*\approx 3\times 10^{-4}B^*.$
 Concerning the AMM it is good to have as reference that the sum of
 the proton and neutron contributions for the same spin projection is
 $|\kappa_p+\kappa_n|\mu_NB\approx 2\times 10^{-5}B^*$ MeV. For
 $B^*=10^3$, its effect is only seen for the smallest temperature we
 consider, and for $B^*=10^4$ its effect is washed out for $T\gtrsim
 100$ keV. Due to the computational effort we will only discuss the
 effect of AMM in one example when its effect is still clearly seen.

\begin{figure*}[!t]
  \begin{tabular}{ccc}
\includegraphics[width=0.33\linewidth]{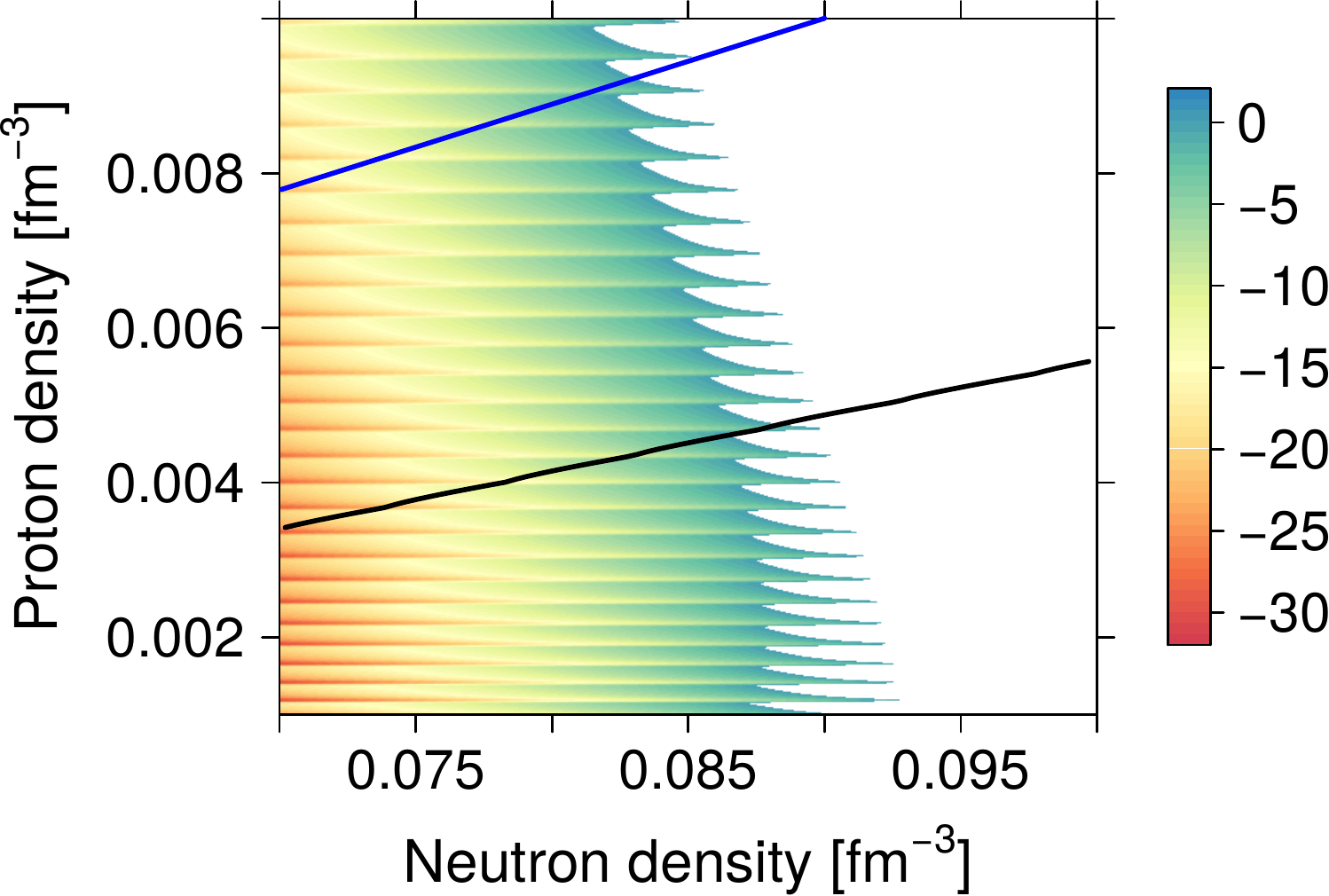} &

\includegraphics[width=0.33\linewidth]{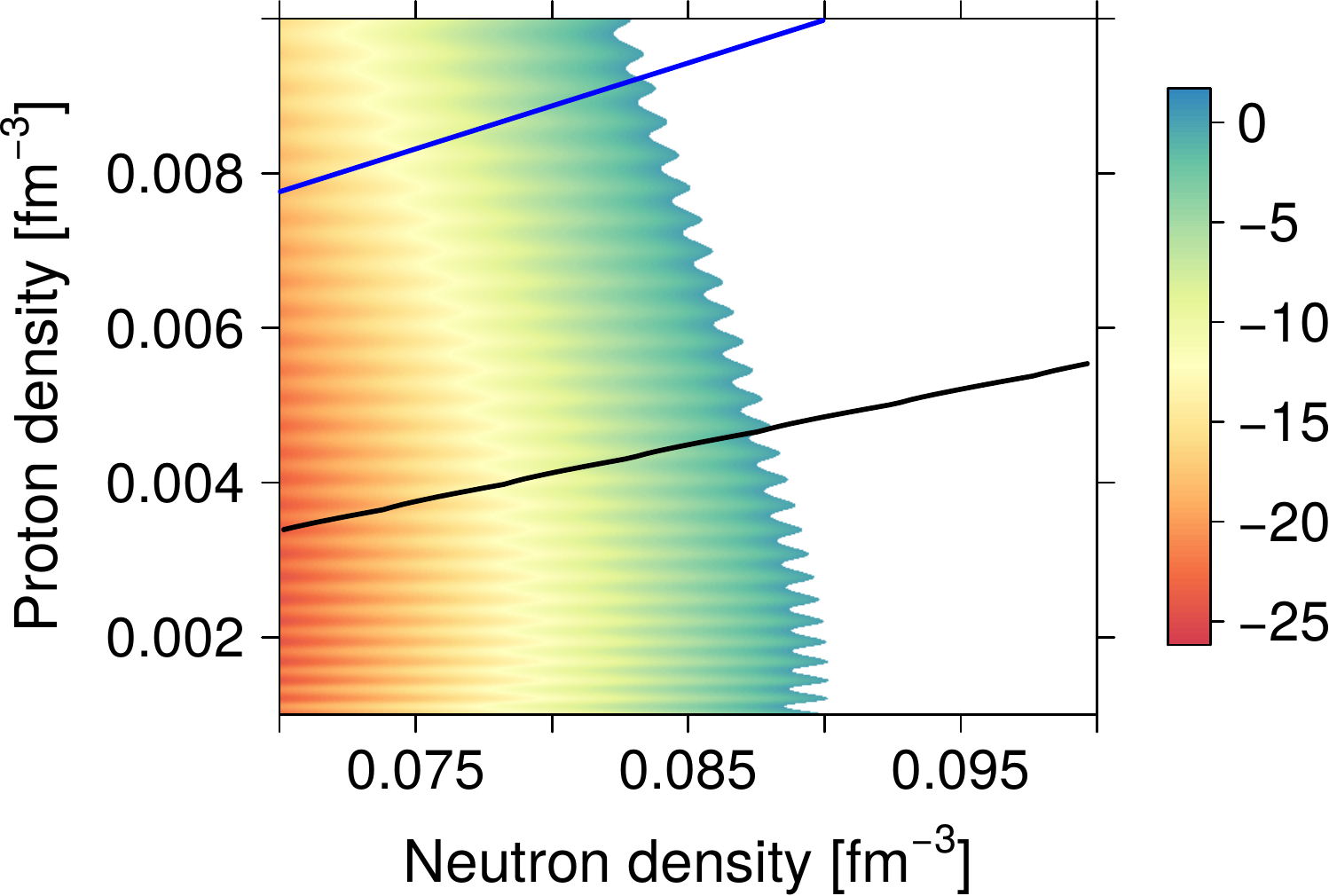} &

\includegraphics[width=0.33\linewidth]{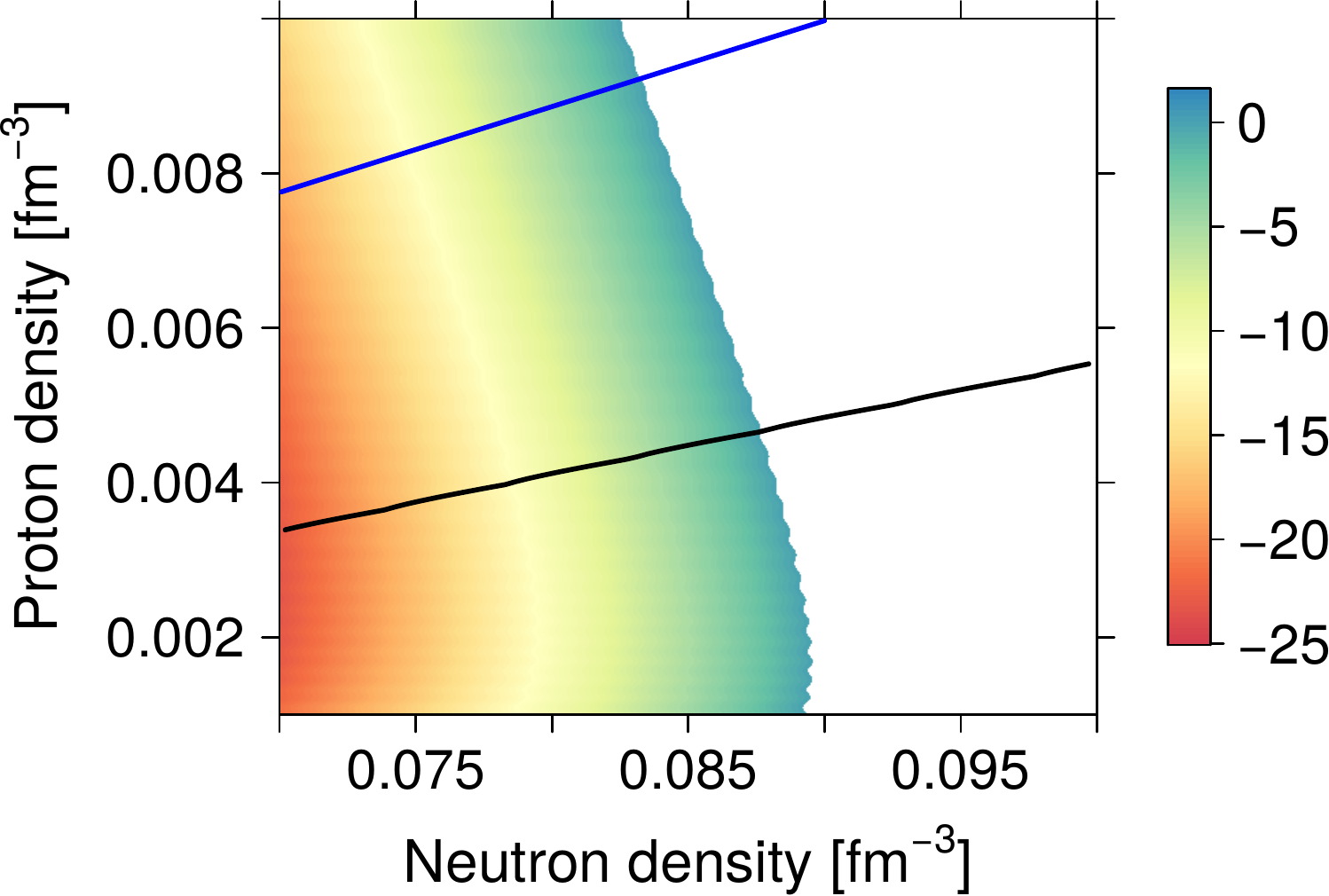} 
    \end{tabular}
\caption{Spinodal sections obtained within the TM1e model for
  $B^{*}=10^{3}$ and temperatures [keV]: 10 (left), 50 (center), and
  100 (right). The colours show the
magnitude of $\lambda_-$ (see Eq. (\ref{lambd}) in units of $(\hbar
c)^3/M^2$. The two lines represented define the $Y_p=0.1$ (top) and
$\beta$-equilibrium (bottom) proton fractions.}
\label{b1d3}
\end{figure*}

In Fig. \ref{T0} we represent, for the TM1e model, the region in the $\rho_p-\rho_n$
space where the eigenvector $\lambda_-$ defined in Eq. (\ref{lambd})
is negative for matter under the effect of a magnetic field intensity
$B^*=10^3$ (left panels) and $B^*=10^4$ (right panels). The colors
indicate the magnitude of the  eigenvector and the lines represent
 the $Y_p=0.1$ line (top) and the $\beta$-equilibrium line (bottom). It is clearly seen the
structure created by the Landau quantization of the proton energies,
which divides the spinodal region into bands which extend well into
regions that are stable for $B=0$. Both sections were obtained without
the inclusion of the anomalous magnetic field (AMM). The main effect
of including the AMM is to divide each band into two narrower bands
that are associated with a different spin projection with respect to
the magnetic field direction. The two bottom panels show details for a
proton density below  0.01 fm$^{-3}$.
This allows to study the transition to non-homogeneous matter with a temperature below 0.5
MeV, in particular, considering a proton fraction below 0.1, as in
protoneutron stars in a neutrino free stage, and temperatures below 1
MeV. The extension of the spinodal bands beyond the $B=0$ curve occurs
with quite shallow minima and, as will be shown below, the temperature
will dissolve these regions. For $T=0$ the $\beta$-equilibrium curves
cross several bands so that between the density when the curve first
crosses the spinodal surface and the density the curve definitely
enters the core, the $\beta$-equilibrium curve has passed through
two unstable  regions. Taking the curve $Y_p=0.1$ three disconnected
unstable regions would occur before the onset of the core.

For the
largest field, $B^*=10^4$, it is clearly
seen that the opening of a new Landau level occurs at defined
values of the proton density and that  the larger the proton density, the
larger the widths of the bands. For this field both proton fractions
represented would give rise to two disconnected unstable regions below
the core.

\begin{figure*}[!t]
  \begin{tabular}{ccc}
\includegraphics[width=0.33\linewidth]{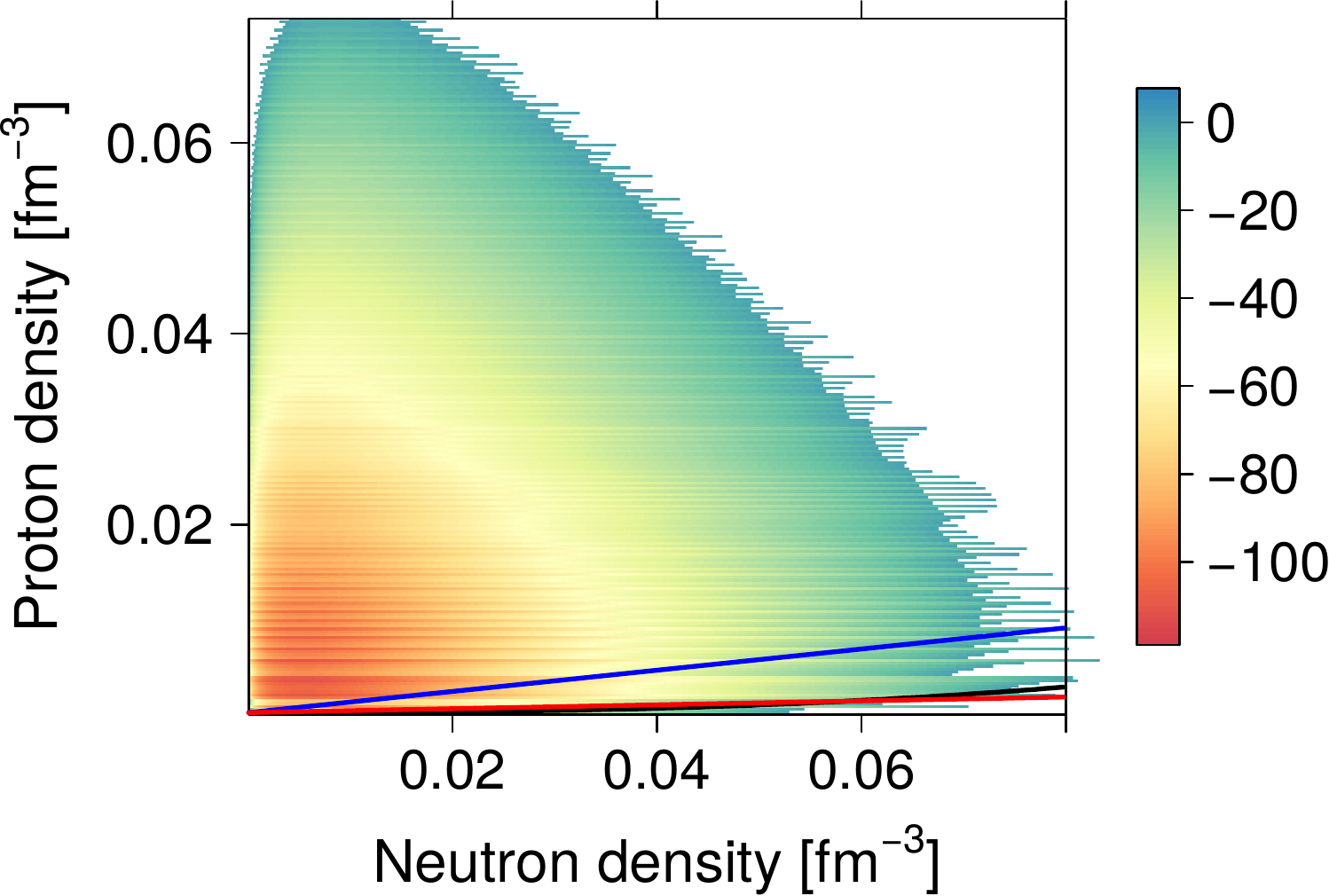} &

\includegraphics[width=0.33\linewidth]{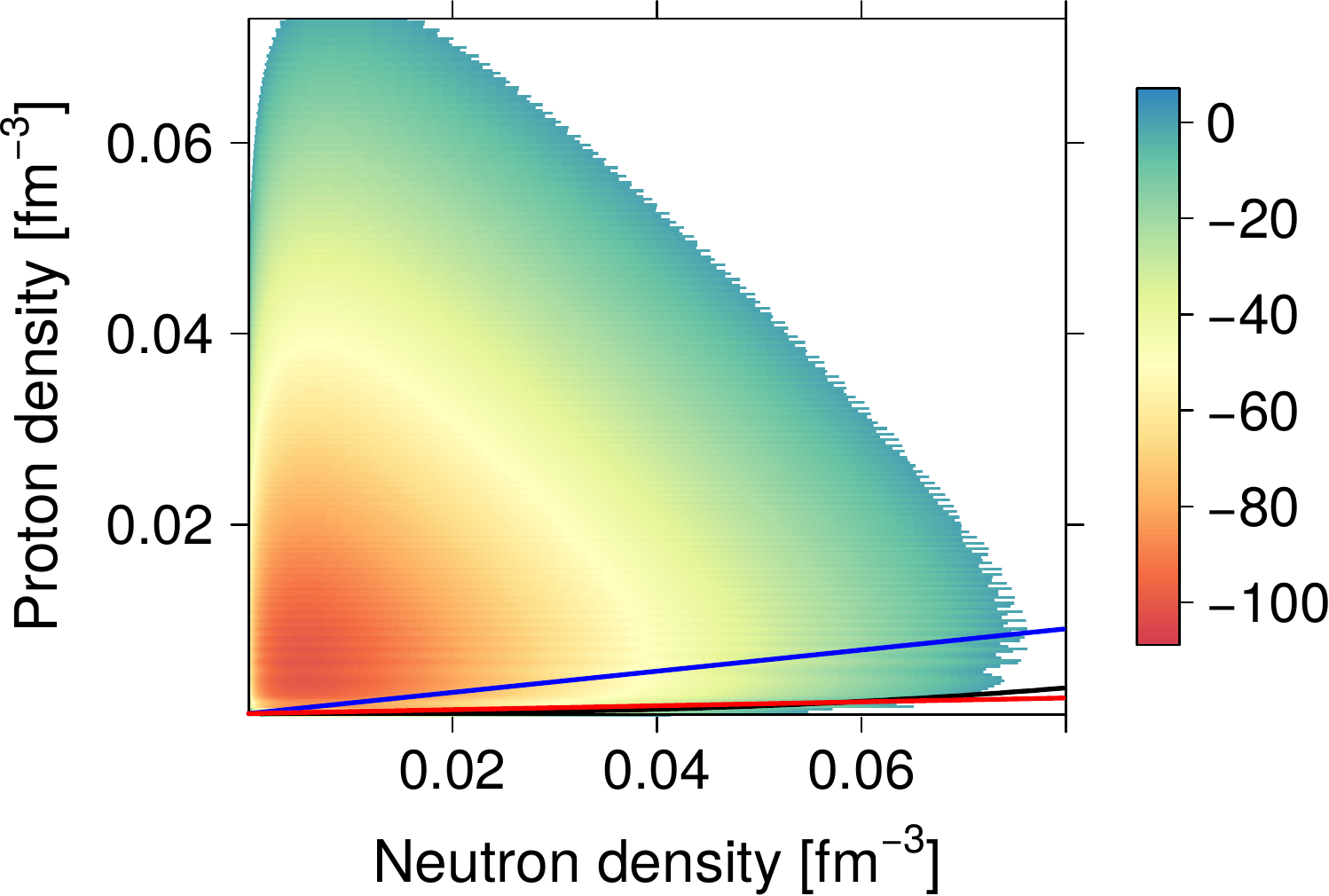} &

\includegraphics[width=0.33\linewidth]{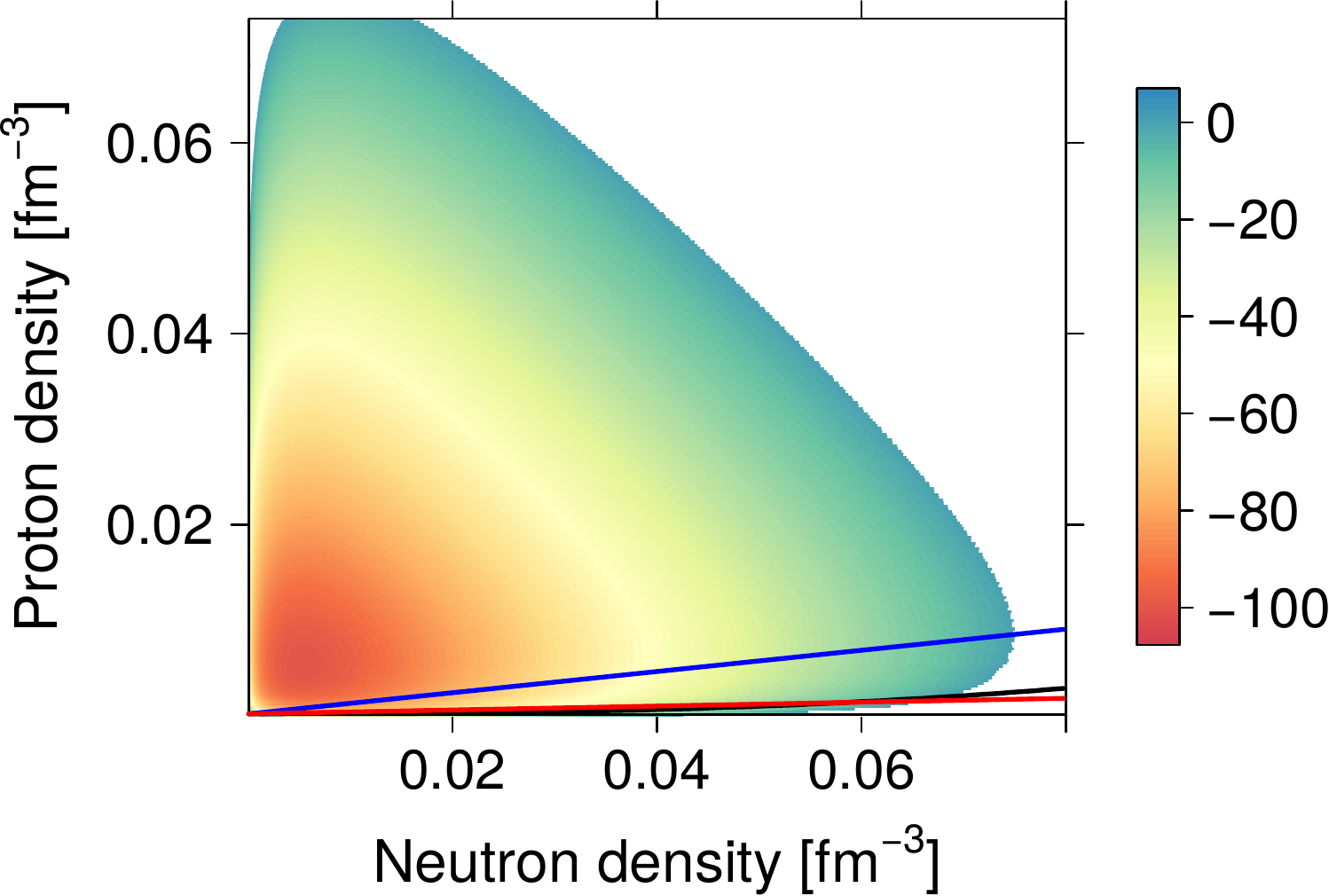} \\

\includegraphics[width=0.33\linewidth]{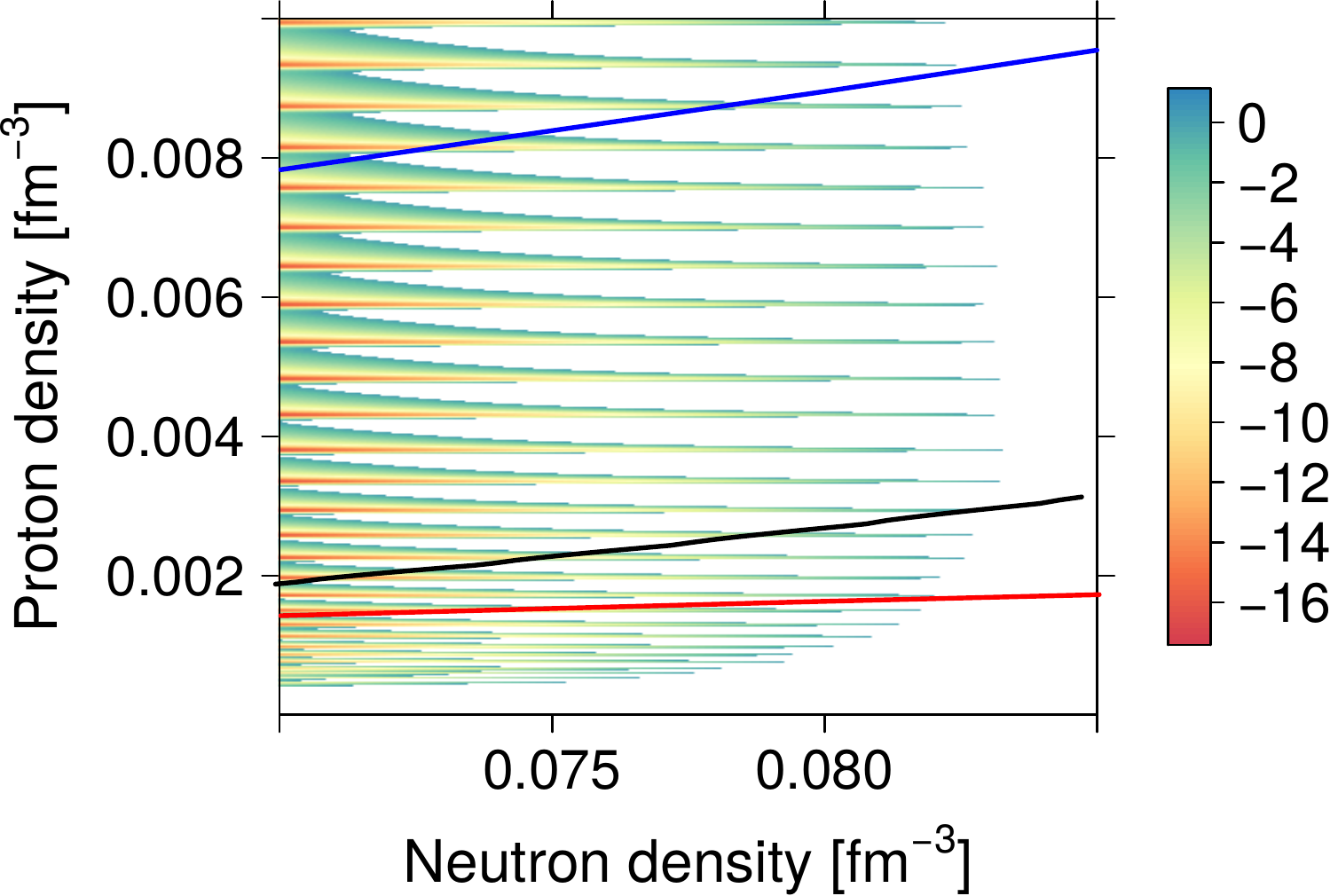} &

\includegraphics[width=0.33\linewidth]{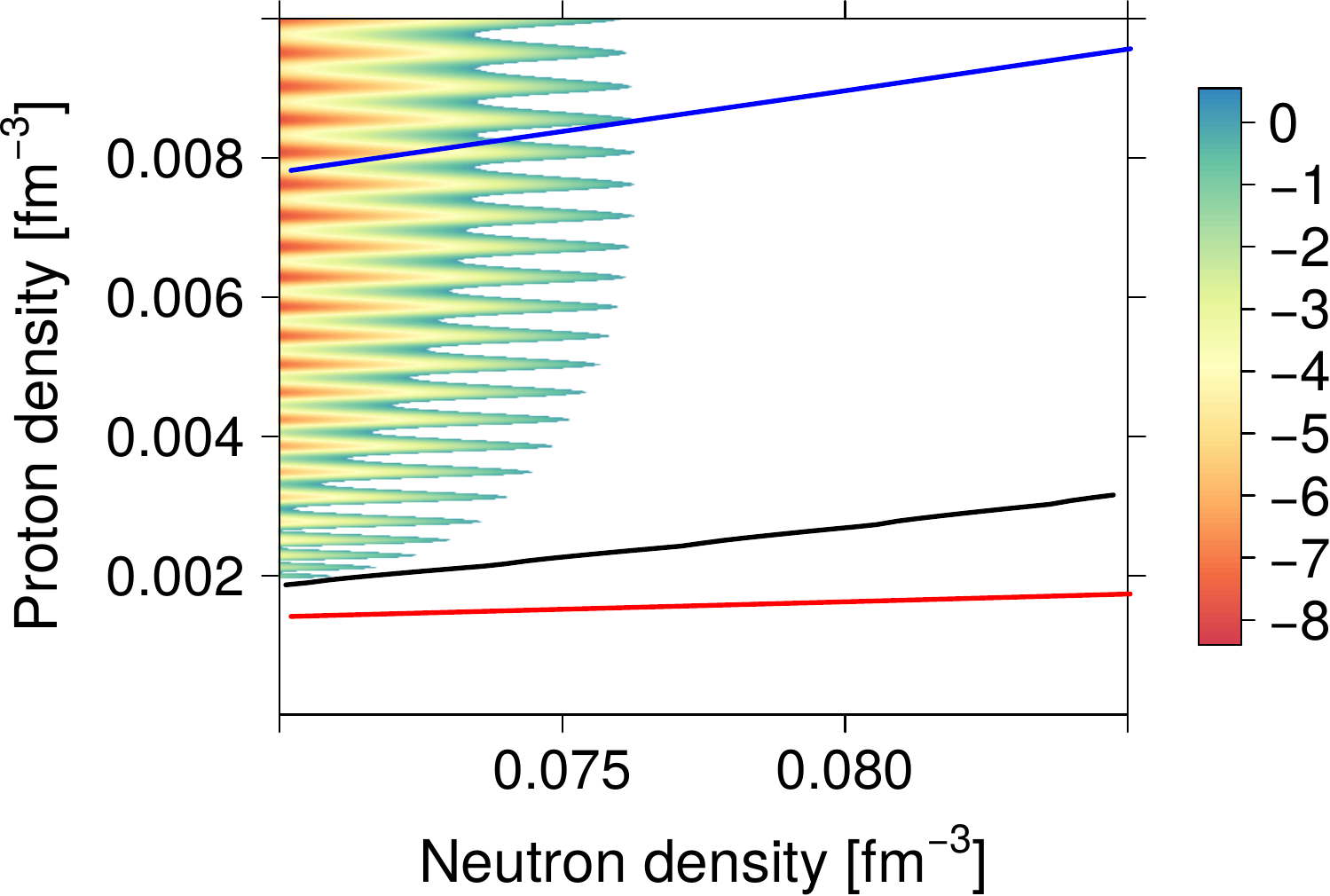} &

\includegraphics[width=0.33\linewidth]{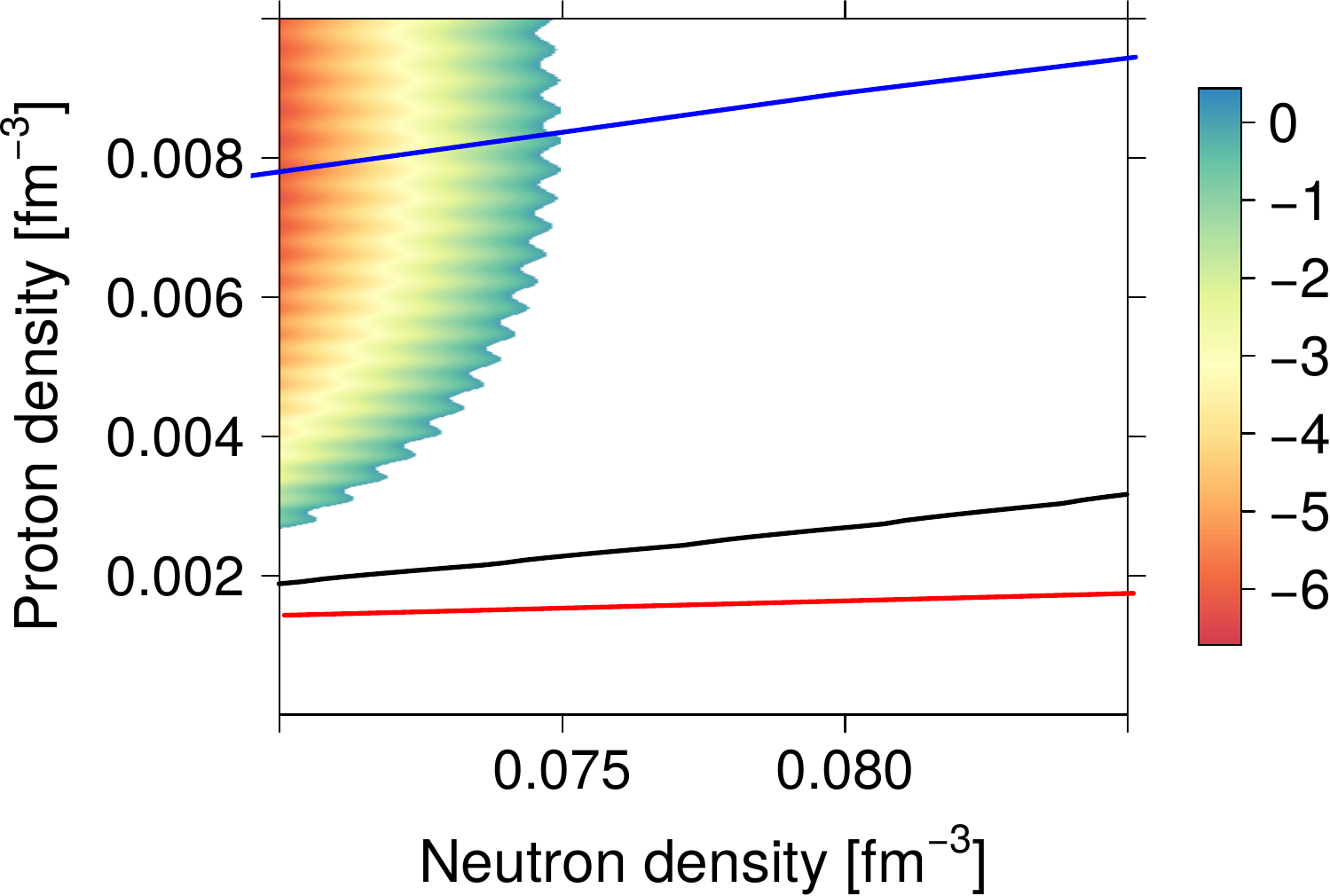}
    \end{tabular}
\caption{Spinodal sections obtained within the TM1 model 
for $B^{*}=10^{3}$ and temperatures [keV]: 10 (left), 50 (center), and
100 (right). The bottom panels contain a detail at the region of
interest for the crust-core transition. The colours show the
magnitude of $\lambda_-$ (see Eq. (\ref{lambd}) in units of $(\hbar
c)^3/M^2$. The three lines represented define the $Y_p=0.1$ (top) and
$\beta$-equilibrium (middle) and $Y_p=0.02$ (bottom) proton fractions. }
\label{tm1b1d3}
\end{figure*}

\begin{figure*}[!h]
  \begin{tabular}{cccc}
\includegraphics[width=0.24\linewidth]{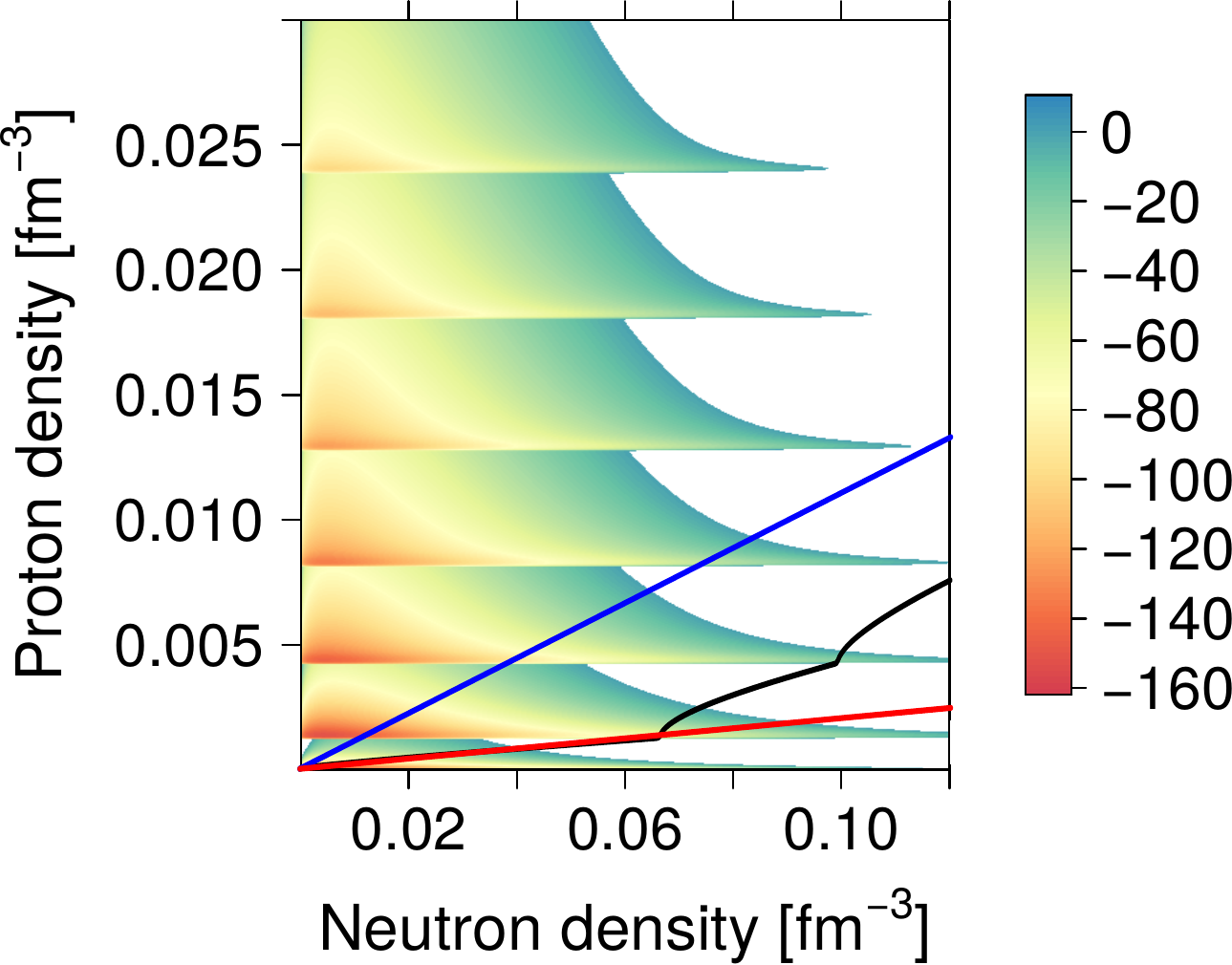} &

\includegraphics[width=0.24\linewidth]{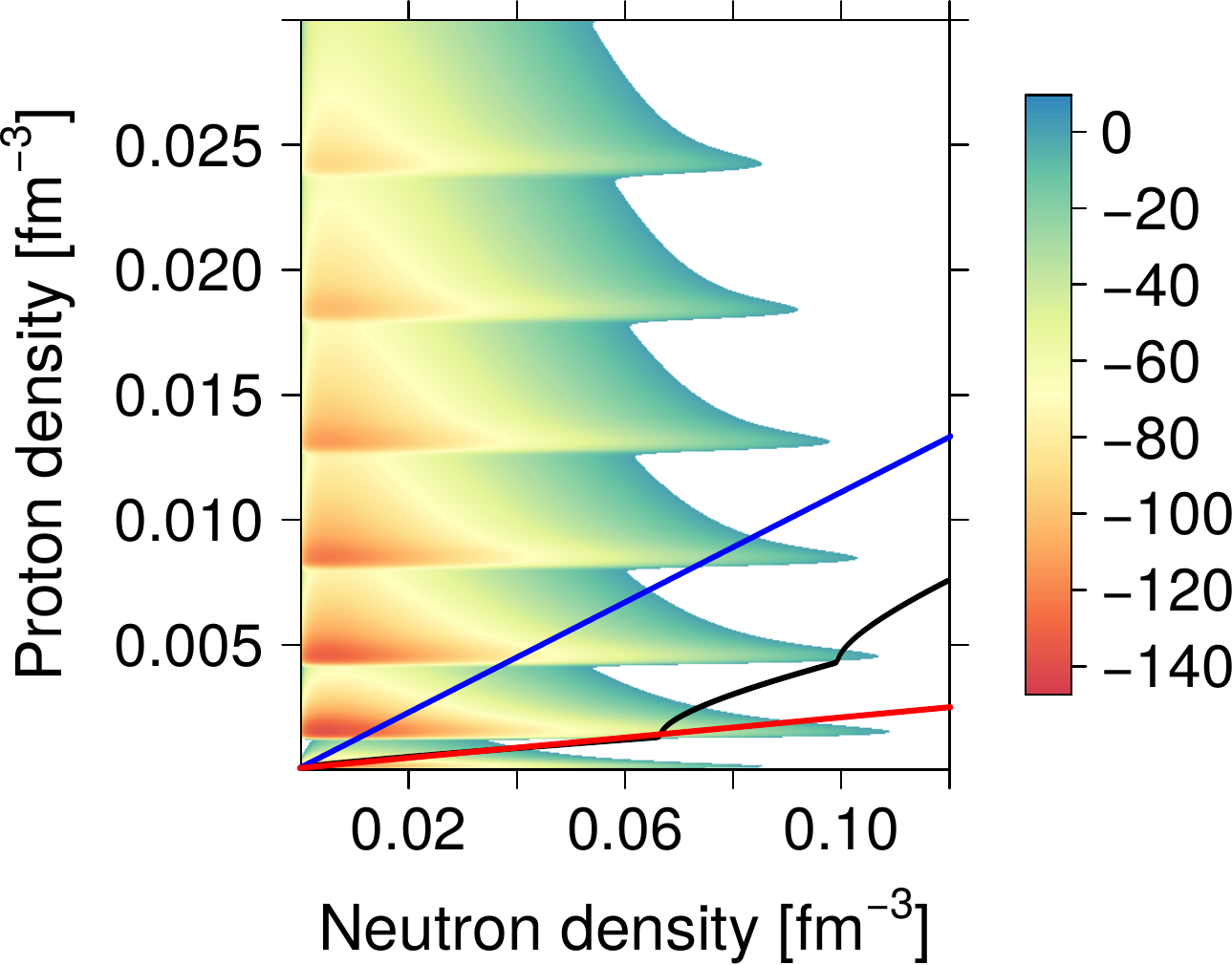} &

\includegraphics[width=0.24\linewidth]{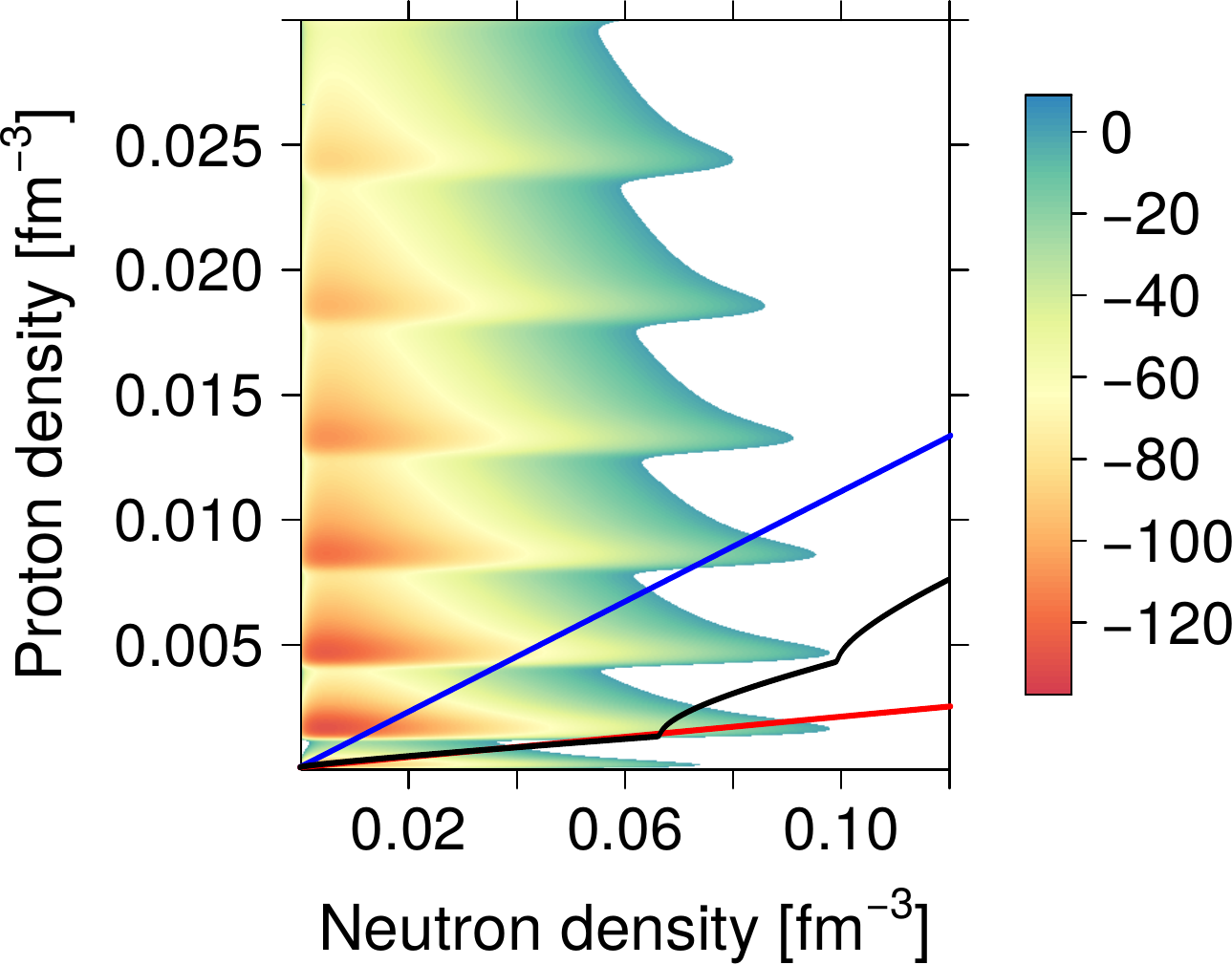} &

\includegraphics[width=0.24\linewidth]{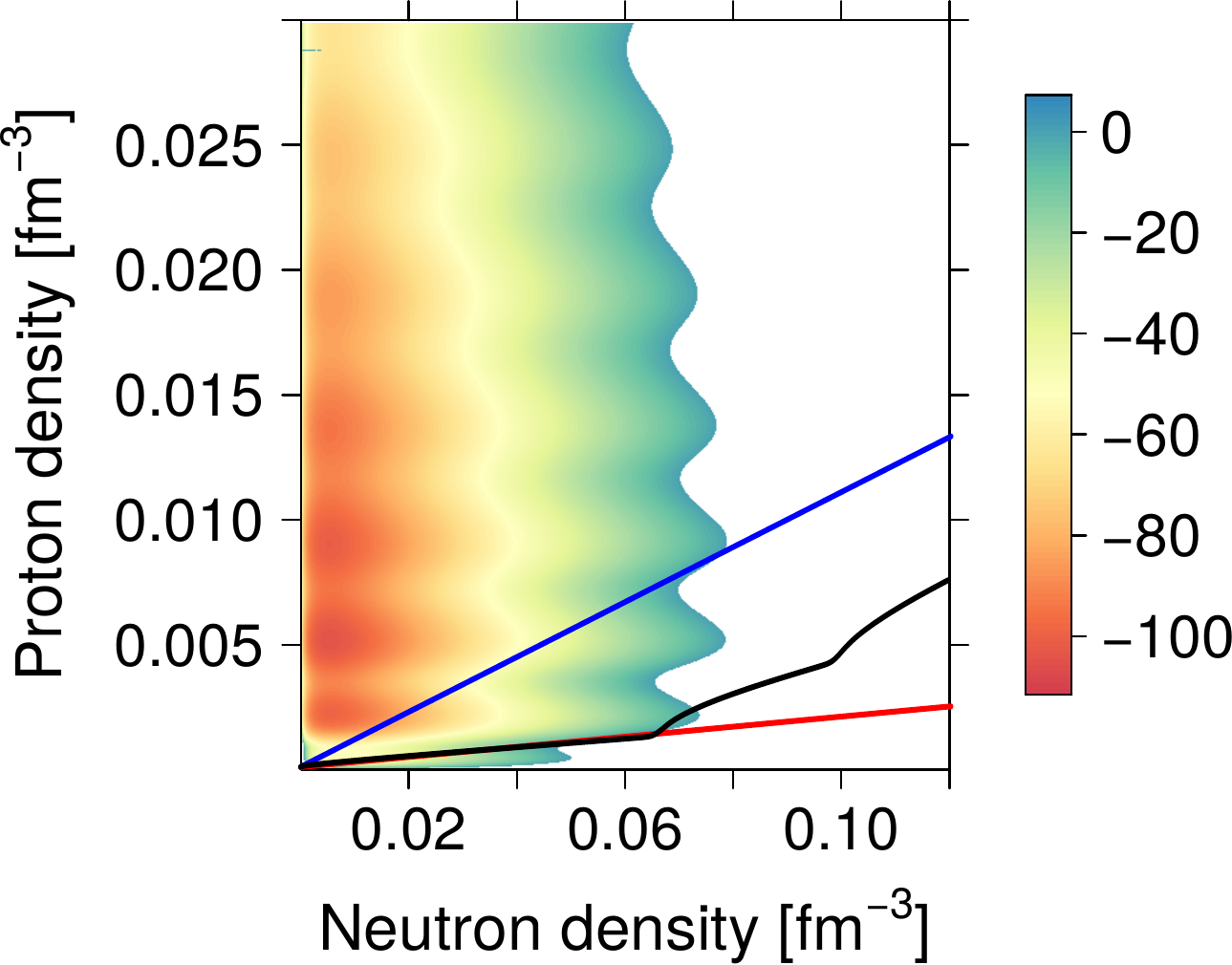} 
    \end{tabular}
\caption{Spinodal sections obtained within the TM1 model 
for $B^{*}=10^{4}$ and temperatures [keV]: 10, 50, 100, and 500 (from
left to right). The colours show the
magnitude of $\lambda_-$ (see Eq. (\ref{lambd}) in units of $(\hbar
c)^3/M^2$. The colours show the
magnitude of $\lambda_-$ (see Eq. (\ref{lambd}) in units of $(\hbar
c)^3/M^2$. The three lines represented define the $Y_p=0.1$ (top) and
$\beta$-equilibrium (middle) and $Y_p=0.02$ (bottom) proton fractions.}
\label{tm1b1d4}
\end{figure*}

We next study the effect of temperature on the spinodal section of the
TM1e model for matter under the effect of a magnetic field with
intensity $B^*=10^3$, see Figs. \ref{b1d3} where the spinodal was
calculated for $T=10$ keV (left) 50 keV (middle) and 100 keV (right) and compare with
Fig. \ref{T0} left panel calculated for $T=0$. Only the detail of the
spinodal necessary to determine the crust-core transition is shown. For $T=10$ keV the
structure of bands is still clearly seen in the interior of the
spinodal surface but the extension of the
bands to regions beyond the $B=0$ spinodal surface is now much smaller:  from the $T=0$  more than 0.01 fm$^{-3}$ length to less than 0.005
fm$^{-3}$. This could be expected since the $\lambda_-$ although
negative took values very close to zero at the extreme ends of the bands.  At $T=50$ keV the band effect is much weaker and at
$T=100$ keV the interior band structure has completely disappeared but
the spinodal surface still shows a slight magnetic field effect. This is
 expectable since the energy interval between two Landau levels is of
 the order of $\lesssim eB/M^*$, where the approximate refers to the low
 magnetic field intensities. For $B^*=10^3$, we have $eB/M^*\approx 0.5$ MeV.
The two curves shown define the $Y_p=0.01$ (top) $\beta$-equilibrium
(bottom) proton fractions. The crossing of these curves with the
spinodal surface indicate that only for the $\beta$-equilibrium proton
fraction two disconnected non-homogeneous regions would
appear. Nonetheless, for the two lowest temperatures $T=10$ and 50 keV
the crust-core transition still occurs at a larger density than predicted
for $B=0$. These three figures also allow to point out a further
detail concerning the structure of unstable region: it is clearly seen
for $T=100$ keV that once the band structure disappears the eigenvalue
$\lambda_-$ is more negative the smaller the neutron density, this is
not the case for the other two cases, and for the $T=0$ spinodals
shown in Fig. \ref{T0}: the bands correspond to more negative values
of $\lambda_-$ at the bottom of the band when the Landau level starts
being filled and much less at the top of the band.

\begin{figure*}[!t]
\begin{tabular}{cc}
\includegraphics[width=0.4\linewidth]{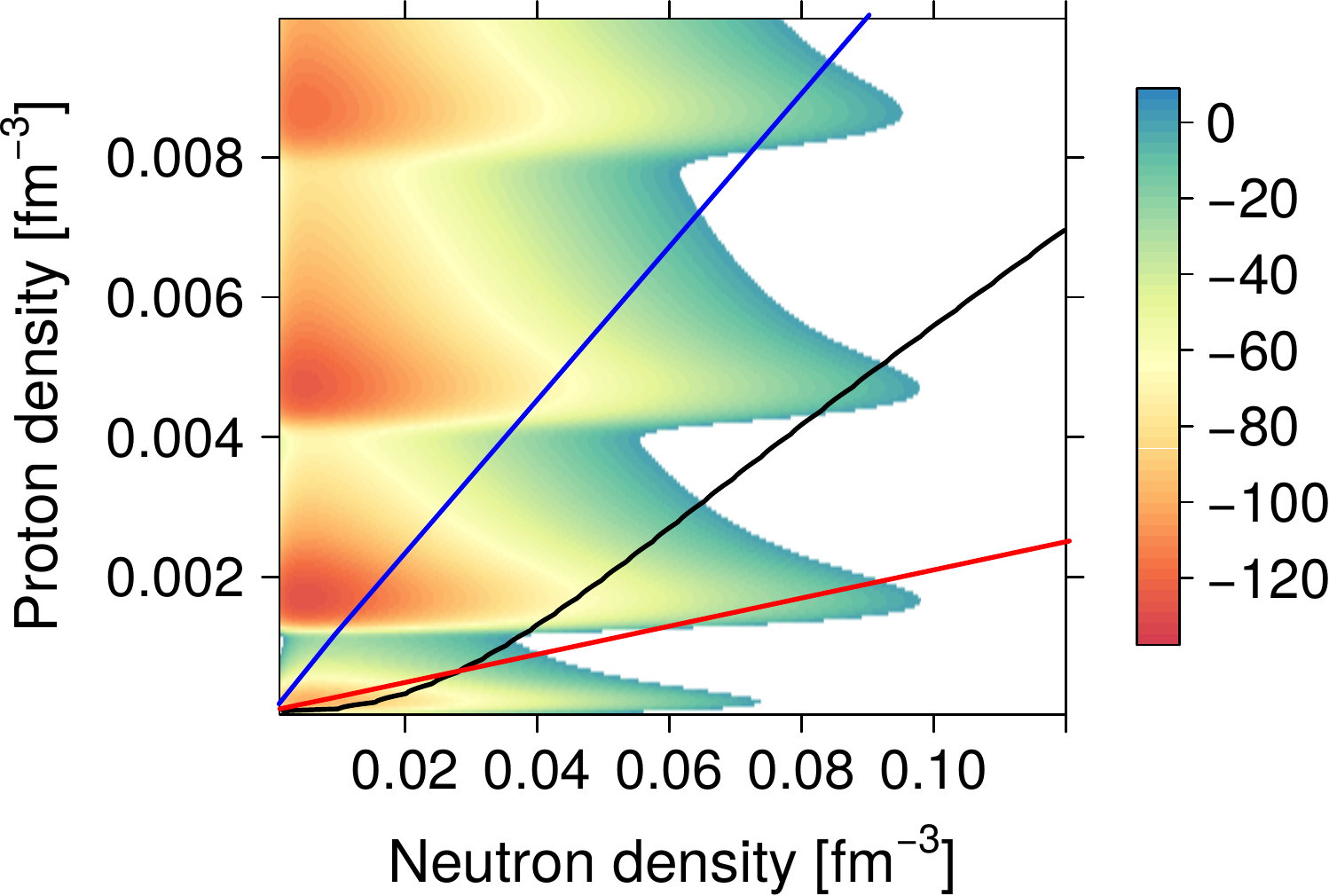} 

\includegraphics[width=0.4\linewidth]{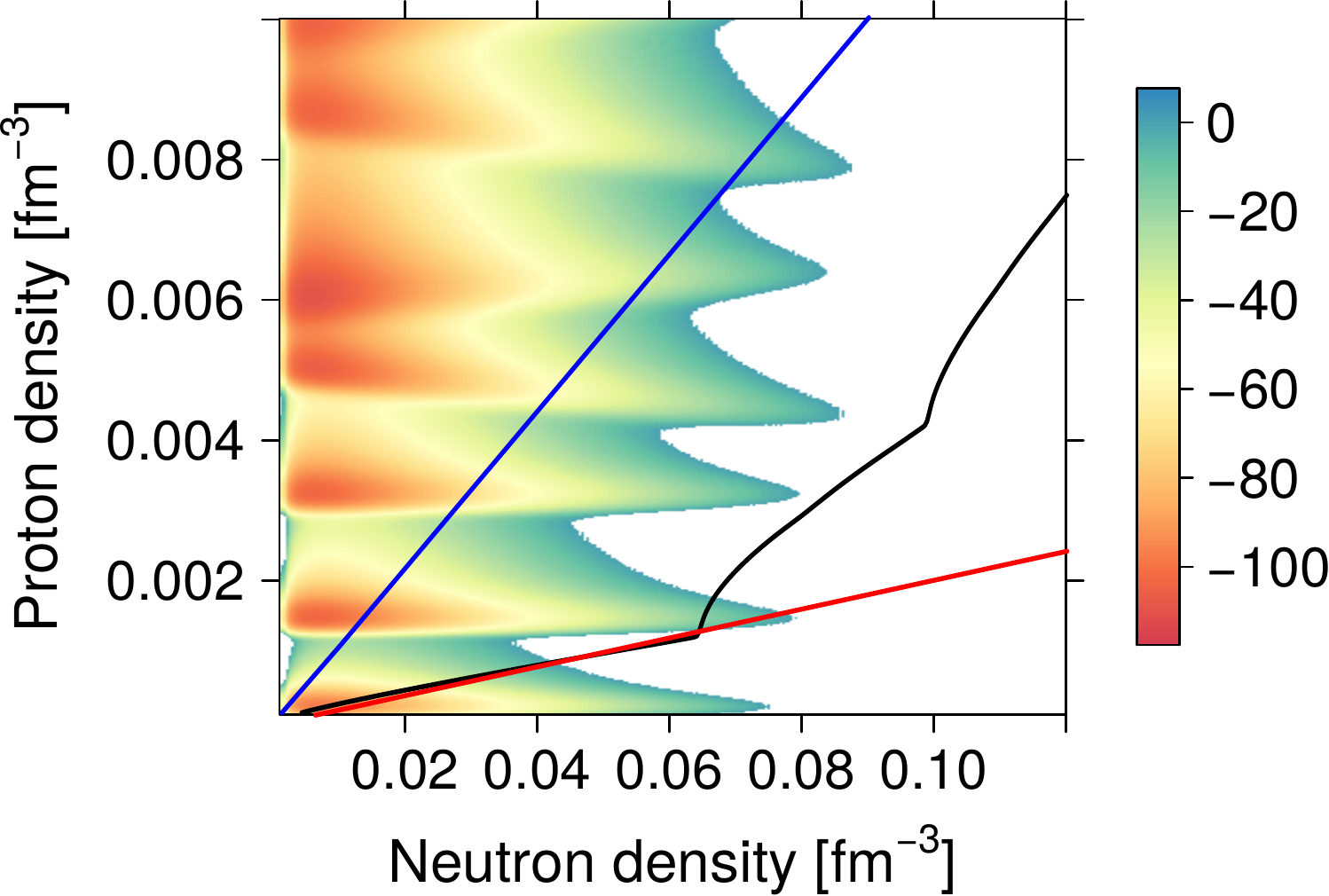}  
\end{tabular}
\caption{Spinodal section obtained within the TM1 model 
for $B^{*}=10^{4}$ and $T=100$ keV without (left) and with (right)
AMM. The colours show the
magnitude of $\lambda_-$ (see Eq. (\ref{lambd}) in units of $(\hbar
c)^3/M^2$. The three lines represented define the $Y_p=0.1$ (top) and
$\beta$-equilibrium (middle) and $Y_p=0.02$ (bottom) proton fractions.}
\label{b1d4amm}
\end{figure*}

 A similar effect is observed for the TM1 model in Figs. \ref{tm1b1d3}:
 while at $T=10$ and 50 keV the band structure is clearly seen, the effect
 of the magnetic field has been washed out for $T=100$ keV, except for
 a small effect at the surface. Comparing
 with TM1e, the TM1 spinodal surface extends to smaller densities. As
 already discussed in other works, see for instance \cite{Ducoin2011},
 this is related to the density dependence of the symmetry energy. However, the extension of the spinodal bands due to Landau
 quantization beyond the $B=0$ surface is larger as discussed in
 \cite{Fang2017}, \cite{Fang2017a} where different models with similar properties were
 discussed. This is due to the fact that at low densities the symmetry
 energy of this model is smaller and allows that asymmetric
 nuclear matter is more bound than matter described by TM1e.
Note, however, that the eigenvalue $\lambda_-$ takes values very close
to zero, and, therefore, temperature dissolves easily these
structures. At $T=10$ keV the $\beta$-equilibrium curve defines five
non-homogeneous disconnected regions while the $Y_p=0.1$ curve defines
three. The $Y_p=0.02$ curve deviates from the $\beta$-equilibrium
curve at the $B=0$ transition density and under-estimates the proton
fraction. However, the qualitative behavior predicted is similar to
the one by  the $\beta$-equilibrium
curve. At $T=100$ keV, there are no disconnected unstable regions
for all scenarios and for $T=50$ keV, there is at most one
disconnected region. We conclude that within TM1, the crust structure is
very sensitive to the temperature evolution, more than within TM1e.

In the following we discuss the effect of a $B^*=10^4$ magnetic field
on the crust-core transition of matter described within the TM1
model. Large fields similar to this one have been reported in studies
which include besides a possible poloidal field also a toroidal field
\cite{Frieben2012,Uryu2014,Uryu2019}.   The sections of the spinodal
surfaces of interest to study neutron rich matter are represented in
Fig. \ref{tm1b1d4} for $T=10, \, 50,\, 100$ and 500 keV. For this
strong field even the largest temperature we consider is not enough to
completely wash out the band structure, although a much weaker effect
is obtained for $T=500$ keV. For the three smallest temperatures the band
structure is similar and only the extension of the bands for larger
neutron fractions is affected with smaller densities being attained
with larger temperatures. This has implications in the structure of the
crust: for $T=10$ keV we identify for $\beta$-equilibrium matter at
least three disconnected non-homogeneous regions; for $T=50$ keV we
still have three disconnected regions, but this number reduces to two
or even only one increasing the temperature to 100 or 500 keV.

In Fig. \ref{b1d4amm} the effect of the AMM is clearly shown in a
scenario when it is not negligible: a magnetic field with an intensity
$B^*=10^4$ is considered and a temperature equal to 100 keV. For such
a strong magnetic field,  the temperature  $T=100$ keV does not remove
the effect of Landau quantization.  As discussed before, this
temperature just reduces the extension of the spinodal bands with
respect to colder scenarios. The
introduction of AMM has as two main effects: i) the splitting of the spinodal
bands into two due to spin polarization. The first band does not
suffer any splitting because it is defined by a single spin
polarization, but all the other bands have been splitted into two, the
separation being larger for the lower bands; ii) the reduction of the
extension of the spinodal bands. Being the effect of the AMM one order
of magnitude smaller than the effect of Landau quantization, its
effect washes away for temperatures that  are approximately one order
of magnitude smaller.

We have included three curves that describe from top to bottom matter
with a proton fraction $Y_p=0.1$, matter in $\beta$-equilibrium ad
matter with  a proton fraction $Y_p=0.02$. This last  fraction was
considered because it represents the proton fraction of the
non-homogeneous inner crust matter close to the crust-core
transition for $B=0$. This value has been considered in previous
studies as in \cite{Fang16,Fang2017}. For the three scenarios described, the crust-core transition is
identified with the crossing of the proton fraction curves with the
spinodal surface. Some conclusions are in order: i) For
$\beta$-equilibrium matter there exists a first crossing from unstable
to stable matter around $0.06$ fm$^{-3}$, however, for $\rho\approx
0.08$ fm$^{-3}$ the curve crosses again an unstable region before
entering definitely the core. Under these conditions the star would
have two disconnected non-homogeneous regions, as already predicted in
\cite{Fang16,Fang2017}. The qualitative conclusion is not different from the one
drawn with $Y_p=0.02$; ii) a similar situation
will occur also for a scenario when $\beta$-equilibrium was still not
attained and the proton fraction is 0.1, although the crossings occur
at different densities. If the AMM is taken into account we still get
two disconnected non-homogeneous regions for $\beta$-equilibrium
matter and not for $Y_p=0.1$ because the extension of the bands is
smaller. If a larger temperature had been chosen the effect of the AMM
would have mostly disappeared.

\section{Conclusions \label{sec5}}

We have discussed the effect of temperature on the
spinodal surface of magnetized nuclear matter within two RMF models
describing symmetric nuclear matter with the same underlying model,
however  having different density dependences of
the symmetry energy. The two models considered, TM1 \cite{tm1} and TM1e
\cite{tm1e}, satisfy, respectively, the constraints imposed by the
recent measurement of PREX-2 and the neutron matter constraints
obtained within a chiral effective field theoretical calculation
\cite{Hebeler2013}. In addition we have briefly commented on the
effect of  including explicitly the nucleon AMM.  The magnitude of its
effect is approximately one order of magnitude smaller than the effect
due to the Landau quantization of the proton energies, and therefore,
temperature washes away its effect at lower temperatures. The main
effect of the nucleon AMM is to split into two each band in the
spinodal region except for the first one, and to  slightly reduce the
extension of the bands beyond the $B=0$ crust-core transition
density. This splitting may define extra disconnected regions but only
in the $T=0$ limit.

From the knowledge
of the spinodals surfaces  it was
possible to make a crust-core prevision for different proton fraction
scenarios and, in particular, to discuss the possible existence of
disconnected non-homogeneous regions and how temperature removes this structure.
The main conclusion was that the prevision of a finite number of  disconnected unstable
regions obtained with $T=0$ MeV  would reduce to a single one for
100-500 keV depending on the field intensity. This implies that during
the cooling or heating of the star,  solid regions will form in
the interior close to the crust-core transition.
Besides, it was shown that this behavior may be more dramatic if the
model that describes the nuclear matter has a large symmetry energy
slope at saturation density, implying a small symmetry energy at
subsaturation densities. In this case it was shown that many
disconnected unstable regions would appear at zero
temperature. However, the free energy density curvature matrix
eigenvalue that defines the instability takes values very close to
zero, and, therefore, temperatures easily melts these structures.
The complexity of
this region and its evolution with cooling may create conditions for a
plastic-flow scenario under very strong magnetic fields and allow for a
field evolution that generates a twisted  corona, as suggested in
\cite{Lander2015,Lander2019}.

\subsection*{Acknowledgments}

This work was partially supported by national funds from FCT (Fundação para a Ciência e a Tecnologia, I.P, Portugal) under the Projects No. UID/\-FIS/\-04564/\-2019, No. UID/\-04564/\-2020, and No. POCI-01-0145-FEDER-029912 with financial support from Science, Technology and Innovation, 
in its FEDER component, and by the FCT/MCTES budget through national funds (OE).




\end{document}